\documentclass[nofootinbib,aps,amssymb,preprint,tightenlines,letterpaper]{revtex4}
\usepackage{graphicx,psfrag}
\newcommand{\G}{\Gamma}
\renewcommand{\d}{\delta}\newcommand{\D}{\Delta}
\newcommand{\e}{\epsilon}
\newcommand{\f}{\phi}\newcommand{\F}{\Phi}
\renewcommand{\j}{\jmath}
\newcommand{\ff}{\varphi}

\renewcommand{\L}{\Lambda}
\renewcommand{\u}{\mu}

\renewcommand{\o}{\omega}\renewcommand{\O}{\Omega}
\renewcommand{\P}{\Psi}

\renewcommand{\S}{\Sigma}
\renewcommand{\t}{\theta}
\newcommand{\X}{\chi}

\newcommand{\da}{\daleth}
\newcommand{\cD}{{\mathcal D}}

\newcommand{\cM}{{ M}}

\newcommand{\cO}{{\mathcal O}}

\newcommand{\cT}{{ T}}

\newcommand{\td}[1]{\tilde{#1}}

\newcommand{\tO}{\td{\cO}}

\newcommand{\be}{\begin{equation}}
\newcommand{\ee}{\end{equation}}
\newcommand{\bea}{\begin{eqnarray}}
\newcommand{\eea}{\end{eqnarray}}

\newcommand{\bra}{\langle}
\newcommand{\ket}{\rangle}

\renewcommand{\le}{\left}
\newcommand{\ri}{\right}

\newcommand{\R}{\mathbb{R}}
\newcommand{\C}{\mathbb{C}}
\newcommand{\N}{\mathbb{N}}
\newcommand{\Z}{\mathbb{Z}}

\newcommand{\SO}{{\mathrm{SO}}}
\newcommand{\SU}{{\mathrm{SU}}}
\newcommand{\SL}{{\mathrm{SL}}}

\newcommand{\su}{{\mathrm{su}}}

\renewcommand{\det}{\mathrm{det}}
\newcommand{\tr}{\mathrm{tr}}

\newcommand{\w}{\wedge}
\newcommand{\co}[2]{\left[#1,#2\right]}

\newcommand{\mm}[4]{\left(\begin{array}{cc}
{#1} & {#2} \\ {#3} & {#4}
\end{array}\right)}

\newtheorem{prop}{Proposition}

\newcommand{\proof}{\textit{Proof : --}\\}

\begin{document}

%**************
%* FRONT PAGE *
%**************
\begin{titlepage}
\title{\large Ponzano-Regge model revisited I:
Gauge fixing, observables and interacting spinning particles}
\author{ Laurent Freidel}
\email{lfreidel@perimeterinstitute.ca}
\affiliation{\vspace{2mm}Perimeter Institute for Theoretical
Physics\\ 35 King street North, Waterloo  N2J-2G9,Ontario,
Canada\\} \affiliation{Laboratoire de Physique, \'Ecole Normale
Sup{\'e}rieure de Lyon \\ 46 all{\'e}e d'Italie, 69364 Lyon Cedex
07, France }

%\vspace{2mm}
\author{David Louapre}
\email{dlouapre@ens-lyon.fr} \affiliation{\vspace{2mm}Laboratoire
de Physique, \'Ecole Normale Sup{\'e}rieure de Lyon \\ 46
all{\'e}e d'Italie, 69364 Lyon Cedex 07, France}\thanks{UMR 5672
du CNRS}

\date{\today}

\begin{abstract}
We show how to properly gauge fix all the symmetries of the
Ponzano-Regge model for 3D quantum gravity. This amounts to do
explicit finite computations for transition amplitudes. We give
the construction of the transition amplitudes in the presence of
interacting quantum spinning particles. We introduce a notion of
operators whose expectation value gives rise to
 either gauge fixing, introduction of time, or insertion of particles,
 according to the choice.
We give the link between the spin foam quantization and the
hamiltonian quantization. We finally show the link between
Ponzano-Regge model and the quantization of Chern-Simons theory
based on the double quantum group of $\SU(2)$.
\end{abstract}

\maketitle

\end{titlepage}
\tableofcontents

\section{Introduction}

It is well known that three dimensional gravity is an integrable system
carrying only a finite number of degrees of freedom \cite{Witten,Deser}.
Despite its apparent simplicity, this theory has not been yet fully solved,
especially at the quantum level.
Such a solution would be relevant to people interested in quantum gravity
since 3 dimensional quantum gravity raises many of the key issues involved in the quantization of gravity
such as: the problem of time, the problem of the construction of physically relevant observables,
the coupling of quantum gravity to particles and  matter fields,
the emergence of semi-classical space-time geometry,
the fate of black holes in a quantum geometry, the effect of cosmological constant,
the holography principle, the quantum causal structure, the role of diffeomorphisms, the sum over topologies...
The notable exception is the absence of gravitons in three dimensions.

The first problem one faces when dealing with the quantum theory
is the variety of techniques and strategies one can use in order to quantize the theory \cite{Carlip1}.
The most prominent ones being the ADM quantization, the Chern-Simons quantization and the spin foam approach.
ADM formulation has the virtue to be expressed in geometrical variables but this quantization scheme
has not been very successful except in a few simple cases \cite{RedPS}.
The Chern-Simons quantization is the most successful quantization scheme so far, which has
allowed  the inclusion of a cosmological constant and particles \cite{ChernS,Carlip:1989nz,schroers}.
The drawback of this approach is that there is a clear lack of geometrical and physical understanding
when working with the Chern-Simons observables. Moreover, this quantization scheme is tailored to three
dimensions and the techniques developed there have no chances to be exported to the higher dimensional case.
Eventually, the spin foam approach is the most promising approach in a direct quantization of
higher dimensional gravity \cite{spinfoam,FKaction}. Being a space time version of loop quantum gravity
it has an interpretation in terms of quantum geometry \cite{QG}. Indeed,
the first model of quantum gravity to be ever proposed by Ponzano and Regge in 1968 was
a spin foam model for Euclidean quantum gravity without cosmological constant
\cite{PR,Ooguri:1991ib}.

Among several strategies, coupling gravity to particles is the simplest and most effective way to
understand the physics and dynamics of quantum spacetime
and the construction of observables.
At the classical level, there is an extensive literature on this subject using a diversity of
related descriptions: the polygonal t'Hooft approach \cite{'tHooft:nj,'tHooft:gz}, the Waelbroeck discrete
phase space approach
\cite{Waelbroeck:dt,Waelbroeck:iz,Waelbroeck:1996sg}, the conformal gauge approach \cite{Cantini:2000dd},
 the Chern-Simons formalism \cite{Vaz:1992dn,Buffenoir:2003zu} and the continuum
first order formalism \cite{Matschull:1997du,Matschull:2001ec}.
Almost all these works concentrate on the case of spinless particles.
At the quantum level, most of the explicit results concerning the scattering of particles have been obtained
using semi-classical techniques \cite{deSousaGerbert:1988bj,deSousaGerbert:1990yp} or the Chern-Simons
formulation \cite{baismuller}.

In our paper, we consider the spin foam quantization of three dimensional gravity
coupled to quantum interacting spinning particles.
We revisit the original Ponzano-Regge model in the light of recent developments
and we propose the first key steps toward  a full understanding of 3d quantum gravity in this context,
especially concerning the issue of symmetries and the inclusion of interacting spinning particles.
The first motivation is to propose a quantization scheme and develop techniques that could
be exported to the quantization of higher dimensional gravity.
As we will see, the inclusion of spinless particles is remarkably simple and natural in this context
and allows us to compute quantum scattering amplitudes.
This approach goes far beyond what was previously done in this context by allowing us to deal
with the interaction of particles.
The inclusion of spinning particles is also achieved. The structure is more complicated but
the operators needed to introduce spinning particles show a clear and beautiful link with
the theory of Feynman diagrams \cite{FLL}.

Another motivation of our work is to give a unified picture of possible quantization schemes.
First, we show that the spin foam quantization is related to the discrete Waelbroeck hamiltonian approach
which we generalize to include spinning particles. This approach is known to be related to 't~Hooft polygonal
approach \cite{Waelbroeck:1996sg} and this gives a first step towards the understanding of scattering amplitude in the 't~Hooft
approach as a spin foam model.
We are aware of an independent work to come \cite{NouiPerez} in the context of loop quantum gravity
leading to results similar to ours.
We also completely unravel in our work, and in a companion paper \cite{mathDL} the link of the Ponzano-Regge model
with the Chern-Simons quantization.
We give a complete treatment of the gauge symmetries of the Ponzano-Regge model generalizing
the work done in \cite{FLdiffeo}.
This opens the way to a finite quantization of Lorentzian gravity in the spin foam approach
\cite{Flor,davids,DLLor}.
We introduce the notion of Ponzano-Regge observables and show that the computation
of the expectation values of these PR-observables leads to gauge fixing, introduction of time or
inclusion of particles, depending on the choice.

We will proceed as follows:
In section (\ref{3dClandQ}), we review three dimensional gravity in the first order formalism, the description
of spinning particles in this context and some facts about path integral quantization and transition amplitudes.
In section (\ref{PR:model}), we review the Ponzano-Regge model, its gauge symmetries and their gauge fixing.
In section(\ref{Oanddual}), we describe the notion of Ponzano-Regge observables,
their application to gauge fixing, or insertion of particles
 and some duality properties
 as well as the link of Ponzano-Regge quantization with Chern-Simons quantization.
 In section (\ref{Hamfor}) we present a hamiltonian description of three dimensional gravity
 allowing the inclusion of spinning particles and  its quantization.
 In section (\ref{tranpartamp}), we present the insertion of interacting
 spinless particles.
In section(\ref{transpinpart}), we describe the inclusion of interacting spinning particles
and the notion of particle graph functional.
In  section (\ref{transitionamp}), we compute explicitly, starting from our general definition
the physical scalar product and the action of he braid group.
We show that we recover what is expected from canonical quantization and
that the braid group action is governed by the quantum group $D(SU(2))$, known as the
kappa deformation of the Poincar\'e group.

\section{3D classical and quantum gravity and particle coupling }\label{3dClandQ}

\subsection{3D classical gravity }
In this part we recall briefly the properties of
3dimensional Euclidean gravity in the first order formalism  \cite{FKaction}.

We consider the first order formalism for 3d gravity. The field
variables are the triad frame field $e_\mu^i$ ($i=1,2,3$) and the
spin connection $\o_\mu^{i}$. The metric is reconstructed as usual
from the triad $g_{\mu \nu}=e_\mu^i \eta_{ij} e_{\nu}^j$ where
$\eta = (+,+,+)$ for Euclidean gravity. In the following,
we will denote by $e^i,
\o^i$ the one-forms $e_\mu^idx^\mu, \o_\mu^i dx^\mu$. We also
introduce the $\SU (2)$ Lie algebra generator $J_i$, taken to be
$i$ times the Pauli matrices,  satisfying $[J_i,J_j] =
-2\epsilon_{ijk}\ \eta^{kl} J_l$, where $\e_{ijk}$ is the
antisymmetric tensor. The trace is such that
$\tr(J_iJ_j)=-{2}\delta_{ij}$. One  defines the Lie
algebra valued one-forms $e=e^i J_i$ and $\o=\o^i J_i$. The action
is
\begin{equation}\label{eqn:BFaction}
S[e,\o]=-\frac{1}{8\pi G} \int_{M} \e_{ijk}\ e^i \w F^{jk}(\o) =
\frac{1}{16\pi G}\int_{M} \tr(e \w F(\o)),
\end{equation}
where $\wedge$ is the antisymmetric product of forms  and
$F(\o)=d\o+ \o \wedge \o$ is the curvature of  $\o$.
The equations
of motion of this theory are
\be
d_\o e =0, \label{eom1} \ \
F(\o)=0 \label{eom2},
\ee
 where $d_\o =d +[\o,\cdot]$ denotes the
covariant derivative.
If $M$ possess some boundaries $\partial M$, the variation of the action
is not zero on on-shell configurations but contains a boundary
contribution
\be\label{boundvar}
\delta S = \frac{1}{16\pi G}\int_{\partial M} \tr(e\wedge \delta{\o}).
\ee
This boundary term vanishes if one fixes the value of the connection
on $\partial M$.
The gauge symmetries of the continuum action
(\ref{eqn:BFaction}) are the Lorentz gauge symmetry
\begin{eqnarray}
\o&\to&g^{-1}dg+g^{-1}\o g, \\ e&\to& g^{-1}eg,
\end{eqnarray}
locally parameterized by a group element $g$, and the translational
symmetry locally parameterized by a Lie algebra element $\f$
\begin{eqnarray}
\o&\to&\o, \\ e&\to& e+d_\o\f.
\end{eqnarray}
and  which holds due to the Bianchi.
identity $d_\o F=0$. This supposes that $\phi=0$ on $\partial M$.
The infinitesimal diffeomorphism symmetry is equivalent on-shell to these symmetry
when we restrict to non-degenerate configurations $det(e)\neq 0$.
The action of an infinitesimal diffeomorphism generated by a vector field
$\xi^\mu$  can be expressed as the combination of an infinitesimal Lorentz
transformation with parameter $\o_{\mu}\xi^\mu$ and a translational
symmetry with parameter $e_{\mu}\xi^\mu$.
The conjugate phase space variables are the pull-back of
$(\o,e)$ on a two dimensional spacelike surface, their Poisson brackets being
\be
\{\o_{\mu}^i,e_{\nu}^j \} = \delta^{ij} \epsilon_{\mu \nu}.
\ee
The generator of the translational gauge symmetry is given by
the pull-back of the curvature on the two dimensional slice,
whereas the pull-back of the torsion generates the Lorentz gauge symmetry.

\subsection{Classical particles}\label{clapar}
We will be interested in the coupling of gravity to particles. For the reader
interested in a more detailed and complete treatment of this problem at the classical
 level we recommend the reference \cite{deSousaGerbert:1990yp} which contains a  comprehensive
 analysis of this problem.
It is well know \cite{Deser} that the metric associated with a spinning
particle of mass $m$ and spin $s$ coupled to 3 dimensional Euclidean gravity is a
spinning cone
\be
ds^2 = (dt +4Gs d\varphi)^2 + dr^2 +(1-4Gm)^2 r^2 d\varphi^2,
\ee
where $m$ is the mass of the particle and $s$ is its spin\footnote{ in $\hbar$ unit  so it has dimension
of an angular momenta.}.
This is a locally flat space, $t$ is the Euclidean time coordinate, $r$ the radial coordinate
measuring at fixed time, the geodesic distance from the location of the particle along constant $\varphi$
geodesics and $\varphi$ is an angular coordinate with the identification $\varphi \rightarrow \varphi +2\pi$.
When $4 G m<1 $ this space can be identified with a portion of Minkowski space.
Lets consider the wedge $0<\varphi < 2\pi (1-4Gm)$, the spinning cone is obtained after
an identification  of the two faces of the wedge by a translation along the $t$ axis of length
$ 8\pi Gs$.
Around $r=0$, which is the location of the particle, there is a deficit
angle of $8\pi G m$ and a time offset of length $ 8\pi Gs$.
The mass of the particle is necessarily bounded by $1/4G$.
A frame field for this geometry
can be given by
\be
e= J_0 dt + (\cos\varphi J_1 + \sin\varphi J_2)dr +
\left( (1-4Gm)r(\cos\varphi J_2 -\sin\varphi J_1) + 4Gs J_0\right)d\varphi,
\ee
and the spin connection by
\be
\o = 2Gm J_0 d\varphi.
\ee
The torsion and the curvature have a distributional contribution at the location of the
 particle\footnote{we use
the distributional identity $dd\varphi =2\pi \delta^2(\vec{x})dx dy$.}
\bea\label{torcur}
d_\o e &=& 8\pi G s J_0 \delta^2(x) d^2x, \\
F(\o) &=& 4\pi G m J_0 \delta^2(x) d^2x,\label{curtor}
\eea
where the delta function is along the plane $t=cste$.
Since the torsion is the generator of Lorentz gauge symmetry we see that
having a spin means that this symmetry is broken at the location of the
particle, also the mass is breaking the translational symmetry at the
location of the particle.
We can explicitly see that this is the case if we perform a Lorentz  transformation
labelled by $g^{-1}$ and then a translational transformation labelled by $-\phi$,
 the equations (\ref{torcur}) then  become
\bea
d_\o e &=& 4\pi G  \jmath \delta^2(x) d^2x,\label{tors} \\
F(\o) &=& 4\pi G  p \delta^2(x) d^2x.\label{curv}
\eea
where $\jmath, p$ are the following Lie algebra elements
\bea
p & =&  m gJ_0 g^{-1},\label{p} \\
\jmath &=&  2s gJ_0 g^{-1} - m [  gJ_0 g^{-1}, \phi ].\label{j}
\eea
$p$ is the momenta of the particle and $\jmath$ the total angular momentum,
they satisfy the constraints
\be\label{jpcons}
-\frac{1}{2}\tr p^2 = m^2; \; -\frac{1}{2}\tr (pj) = 2ms.
\ee
 From the canonical point of view these constraints are first class \cite{deSousaGerbert:1990yp},
 the mass constraint generates time reparameterization and the spin constraints
 generates $U(1)$ gauge transformation $g\rightarrow gh$.
 Due to the breaking of symmetry at the location of the particle
 the gauge degrees of freedom $g,\phi$ become dynamical
 (modulo the remnant reparameterization plus $U(1)$ gauge symmetry):
 $g$ describes the Lorentz frame of the particle,
 $\phi$ describes the position of the particle.
 Moreover the knowledge of $p,\jmath$ is enough to reconstruct  $g,\phi$ modulo the remnant gauge symmetries.
 Indeed $gH$ is determined by (\ref{p}).
If we denote by $x^a_\bot$ the position of the particle perpendicular to the momenta,
$\phi = \frac{(p\cdot \phi)}{m^2}p + x_\bot $, then
$ x_\bot = \frac{1}{m^2} [\jmath,p]$, also $\jmath^2 = s^2 + m^2 x_\bot^2$.

We can easily understand the canonical commutation relations of $p,\jmath$ from the
equations (\ref{tors},\ref{curv}). Since the LHS of (\ref{tors})
is the generator of Lorentz transformations and the LHS of (\ref{curv}) is the
generator of translational symmetry,
these constraints are first class and from their canonical algebra we can
easily deduce that the Poisson algebra of $p,\jmath$ is given by
\be \label{jpcom}
\{\jmath^a,\jmath^b\}= -2 \e^{abc}\jmath_c, \, \{\jmath^a,p^b\}=-2 \e^{abc}p_c,\, \{p^a,p^b\}=0.
\ee

This analysis shows that, instead of treating the gravity degrees of freedom
and the particle degrees of freedom as separate entities,
we can reverse the logic and consider that the equations
 (\ref{torcur},\ref{curtor}) are defining equations for a spinning particle.
This allows us to describe a particle as a singular configuration of the gravitational field
giving a realization of matter from geometry.
The `would-be gauge' degree of freedom \cite{Carlip} are promoted to dynamical degrees
of freedom at the location of the particle. This is the point of
view we are going to take in this paper.
In order to get the equations (\ref{torcur},\ref{curtor}) from an action principle
we have to add to the gravity action (\ref{eqn:BFaction})
the following terms
\be\label{partact}
\overline{S}_{P_{m,s}}(e,\o)= -\frac{1}{2} \int dt\ \tr[(me_t+2s\o_t)J_0],
\ee
where the integral is along the worldline of the particle.
This action describes a `frozen' particle without degrees of freedom.
We have seen in (\ref{tors},\ref{curv}) that the particle degrees of freedom
are encoded in the former gauge degrees of freedom. To incorporate the dynamics of the particle
 we perform the transformation
\be
\o \rightarrow \tilde{\o}= g^{-1}\o g + g^{-1}dg , \, \, e \rightarrow \tilde{e}= g^{-1}(e + d_{\o}\phi)g
\ee
the action (\ref{partact}) becomes
\be
\overline{S}_{P_{m,s}}(\tilde{e},\tilde{\o})=
-\frac{1}{2} \int dt\ \tr[e_t p +\o_t \jmath]
+ S_{P_{m,s}}(g,\phi)
\ee where
the first term describes the interaction between the particle degree
of freedom and gravity. The second term
\be
S_{P_{m,s}}(g,\phi)=-\frac{1}{2} m \int dt \tr(g^{-1}\dot{\phi} g J_0) - s
\int dt \tr(g^{-1}\dot{g}J_0),
\ee
is the action for a relativistic spinning particle in a form first describe
by Sousa Gerbert \cite{deSousaGerbert:1990yp}.
One  sees again that the original gauge degree of freedom are now
promoted to dynamical degree of freedom describing the propagation of a particle.

Note that $S_{P_{m,-s}}(g,\phi) = S_{P_{m,s}}(g\e,-\phi)$ where
$$\e =\left(
\begin{array}{cc}
  0 & -1 \\
 1 & 0
\end{array}
\right), $$ so this action describes a particle carrying  both spin $s$ and
$-s$, which is necessary in order to have $P,T$ invariance.

\subsection{Quantum amplitudes} \label{Quamp}

So far we have described the classical features of 3d Euclidean gravity
in the first order formalism. We want now to present some general features
of partition functions, transition amplitudes and particle insertions
which are necessary in order to deal with the explicit
quantification of the theory that we will present in the next sections.
Before going on, one should keep in mind that the theory we are
studying is classically equivalent to usual gravity as long as we
restrict to non degenerate configurations of the frame field,
$det(e)> 0$. This condition of non degeneracy is not easy to
implement at the quantum level and we will pursue the quantification
without worrying about this issue. We have however to keep in mind that
the resulting quantum theory is expected to depend
on this issue as shown by the study of symmetry \cite{Matschull:1999he}
or by the study of spacetime volume expectation value \cite{Freidel:1998ua}.

Let us also emphasize that we are dealing in this paper with the quantification
of Euclidean gravity. Euclidean gravity is a theory which admits an hamiltonian formulation
and its quantification is a well posed problem as shown for instance by the Chern-Simons formulation.
Since this is often a source of confusion let us stress that
the quantification of Euclidean gravity has nothing to do with a Wick rotation of Lorentzian
gravity, which has never been proven to make sense. This is why in the following we are dealing with
quantum amplitudes and not ill-defined, Wick-rotated, statistical amplitudes.

Let us eventually emphasize that the discussion in this section is a formal
discussion which will help us to introduce all the relevant notions
and notations that we will be  properly defined in the subsequent sections.

\subsubsection{Partition function}\label{transamp}
 At the quantum level, the prime object of interest is the partition
function. Given a closed manifold $M$ we consider
\begin{equation}\label{eqn:continuum-PI}
Z_{M}=\int \cD \o \cD e \exp\le[\frac{i}{16\pi G} \int_M \tr\le(e\w F(\o)\ri)\ri].
\end{equation}
In order to have a proper definition of $Z_M$, we should not overcount the
configurations which are equivalent by gauge symmetries and one should
restrict the integral over gauge equivalent classes of fields
$(e,\o)$ by gauge fixing.
We will call `kinematical observable' a general functional of the field ${\cal O}(e,\o)$.
One of the main object of interest at the quantum level is its expectation value
\be
\bra {\cal O} \ket = \int \cD \o \cD e \exp\le[\frac{i}{16 \pi G}\int_M \tr\le(e\w F(\o)\ri)\ri] {\cal O}(e,\o).
\ee
A `physical observable' is a gauge invariant (under Lorentz symmetries and diffeomorphisms)
observable.Of course,
the final interest is in physical observables, but the kinematical observables will also be
of interest to us. For instance, a gauge fixing procedure or the introduction of a particle
is realized by the insertion of a kinematical observable.

\subsubsection{  Transition amplitudes for pure gravity}
In the case of a manifold $M$  with boundary we need to specify
boundary data
in order to have a well defined path integral.
The unconstrained phase space of $2+1$ gravity is given by
conjugate pairs $(\bar{e},\bar{\o})$, which are the pull-back  of
$(e,\o)$ on $\partial M$.
As we have seen in eq.(\ref{boundvar}) the natural choice is
to take the boundary connection $\bar{\o}$ to be fixed in
$\partial M$. This amounts to a choice of polarization where we consider
wave functionals to be functionals of the connection $\P(\bar{\o})$.
Another natural choice, is to fix $\bar{e}$, hence the geometry, on the
boundary. In this case we have to add a boundary term to the action
\be
S_{b}= -\frac{1}{16\pi G}\int_{\partial M} \tr( e\wedge \o).
\ee
The path integral, with the set of boundary data,
defines for every manifold $M$ with a boundary, a wave functional
\be
G_M(\bar{\o}) = \int_{{\o|\partial M} =\bar{\o}} \cD \o \cD  e\ e^{iS[\o,e]}.
\ee
We can choose to split the boundary into  initial and final boundary
components
$\partial M = \Sigma_i \coprod {\Sigma}_f$, where
$\Sigma_f$ is equipped with the orientation coming from
$M$ and $\Sigma_i$ is equipped with  the opposite orientation.
For instance when $M =\Sigma \times I$, we have
$\Sigma_i = \Sigma \times \{0\}$, $\Sigma_f = \Sigma \times \{1\}$.
In this case $ \bar{\o} =(\bar{\o}_f, \bar{\o}_i)$ and
$G_M(\bar{\o}_f, \bar{\o}_i)$ is  the
kernel of the propagator allowing us to compute { transition amplitudes}
between two states
\begin{eqnarray}\label{physprod}
<\F_f|\F_i>=\int \cD \o\cD e \, e^{iS[\o,e]} \F_f^*(\bar{\o}_f)\F_i(\bar{\o}_i)
= \int \cD \bar{\o}\, \F_f^*(\bar{\o}_f)G_M(\bar{\o}_f, \bar{\o}_i)\F_i(\bar{\o}_i).
\end{eqnarray}
This scalar product is interpreted to be the {\it physical} scalar product, hence it
should be positive. It is not, however, a definite positive product: it is expected to
have a kernel characterized by the hamiltonian constraint.

A natural basis for gauge invariant functionals of $\bar{\o}$ is given by spin network
functionals \cite{Rovelli:1995ac,Ashtekar:1994mh}.
Such functionals are constructed from a closed  trivalent graph $\G$ whose edges $\bar{e}$ are
colored by representation $j_{\bar{e}}$ of $\SU(2)$.
We denote such states by
$\F_{(\G,j_{\bar{e}})}(\bar{\o})$.
Such states are constructed by first taking the holonomy of the connection along
the oriented edges $\bar{e}$
$ g_{\bar{e}}(\bar{\o}) = \overrightarrow{\exp}\left(\int_{\bar{e}} \bar{\o}\right)$,
then taking the matrix elements of this object
$D_{j_{\bar{e}}}(g_{\bar{e}}(\bar{\o}))$ in the spin $j_{\bar{e}}$ representation
and finally by contracting
$\otimes_{\bar{e}} D_{j_{\bar{e}}}(g_{\bar{e}}(\bar{\o})) \in
\otimes_{\bar{e}} V_{j_{\bar{e}}}\otimes \bar{V}_{j_{\bar{e}}}$ using
invariant and normalized tensors $C: V_1\otimes V_2 \otimes V_3 \rightarrow \C$ at the vertices $\bar{v}$ of
$\Gamma$ \footnote{the space of trivalent intertwiners is one dimensional in the case of $\SU(2)$,
since we consider only trivalent graph we do not need to specify an intertwiner label at vertices.}. So
\be
\F_{(\G,j_{\bar{e}})}(\bar{\o}) =
\bra \otimes_{\bar{v}} C_{\bar{v}} |\otimes_{\bar{e}} D_{j_{\bar{e}}}(g_{\bar{e}}(\bar{\o}))\ket.
\ee
The construction needs $\G$ to be an oriented graph, however the
results are independent of this orientation in the case of $\SU(2)$.
Given a graph $\Gamma$, we can consider the vector space generated by
all spin networks with support $\Gamma$, i-e the space of all
cylindrical functions  of the connection having the form
$\Phi= \sum_{j_{\bar{e}}} c_{j_{\bar{e}}} \F_{(\G,j_{\bar{e}})} $ where the sum
is a finite sum over admissible coloring.
We denote this vector space by ${\cal H}_{\Gamma}$.
This vector space can be promoted to an Hilbert space with the norm
\be
||\F||^{2}= \int\prod_{\bar{e}}dg_{\bar{e}} |\F(g_{\bar{e}})|^{2},
\ee
$dg$ being the normalized $\SU(2)$ Haar measure.
This Hilbert space admits a basis independent description
\be
{\cal H}_{\Gamma}= L^{2}(G^{|\bar{e}|}/G^{|\bar{v}|}),
\ee
where $|\bar{e}|$ denotes the number of edges of $\Gamma$, $|\bar{v}|$
the number of vertices and the action of $G^{|\bar{v}|}$ on $G^{|\bar{e}|}$ is given by the
structure of the graph: $g_{\bar{e}}\rightarrow h_{t(\bar{e})}^{-1} g_{\bar{e}} h_{s(\bar{e})}$,
$t(\bar{e})$ and $s(\bar{e})$ being the target and the source of the edge $\bar{e}$.

\subsubsection{Transition amplitudes with massive spinning particles}\label{spinningpart}

 In this paper, we also want to describe the case where  particles are  present.
In this case, as we have seen in the previous section, the gauge symmetries are broken
at the location of the particle, moreover the previously gauge degrees of freedom are
now promoted to dynamical degrees of freedom of the particle.
In order to describe the insertion of particles, we have to add the particle action
(\ref{partact}) to the gravity action. This  breaks the initial
gauge symmetry. The propagator is given by
\be\label{noncovint}
G_M(\bar{\o}) = \int_{{\o |\partial M} =\bar{\o}}
\cD \o \cD  e\, e^{iS[\o,e]+\sum_{n} \overline{S}_{P_{n}(\o,e)}}.
\ee
where $n\equiv (m_n,s_n)$ denotes a mass and spin.
We can decompose the path integral into an integral
over the gravitational fields and an integral over the gauge degrees of
freedom at the location of the particles
\be\label{covint}
G_{M,P_n}(\bar{\o},\bar{g}_{n,i},\bar{g}_{n,f}) = \int_{{\o|\partial M} =\bar{\o}} \cD \o \cD  e\
e^{iS[\o,e]} \prod_{n}G_{P}(\o_{P_n},e_{P_n},\bar{g}_{n,i},\bar{g}_{n,f}),
\ee
\be
G_{P_{m,s}}(\o_P,e_P,\bar{g}_{i},\bar{g}_{f})=\int_{{g|\partial M} =(\bar{g}_{i},\bar{g}_{f})}
\cD g \cD \phi\ \exp({\int dt -\frac{1}{2}\tr[e_t p +\o_t \jmath]+{S}_{P_{m,s}}(g,\phi)}).
\ee
$(\o_P,e_P)$ denotes the value of $(\o,e)$ at the location of the particle.
In the integral (\ref{covint}) over $(\o,e)$, the action is gauge invariant,
contrary to the  one in (\ref{noncovint}),
and the integration is understood to be over gauge equivalence classes.
If we concentrate first on the particle integral $G_{P}(\o,e)$ and
suppose for the sake of the argument that $\o=e=0$ along the
particle worldline, then the interaction term disappears and we are
left with ${S}_{P_{m,s}}(g,\phi)$.
The propagator $G_{P}$ is then the propagator of
a relativistic particle and as such an operator acting on the Hilbert space
obtained by the quantification of (\ref{jpcom},\ref{jpcons}).
Such a quantification can be easily described as follows.
We take our kinematical Hilbert space to be $L^{2}(G)$, such an
Hilbert space is spanned by the Wigner functions $D^I_{nk}(g)$ where
the spin $I$ is an half integer and $n,k$ are the representation
indices.
We define the operators $\hat{p}$ to act by multiplication, and
$\hat{\jmath}$ to act as the right invariant derivative
\be
\hat{p}^a D^I_{nk}(g) = m\tr( gJ_0 g^{-1}J^{a})D^I_{nk}(g); \, \,
\hat{\jmath}_{a}¥ D^I_{nk}(g) = -D^I_{nk}(J_{a}g).
\ee
The mass constraint $p^{2}=m^{2}$ is trivially satisfied, the spin
constraint implies that $k=-s$
\be\label{pjs}
\hat{p}^a\hat{\jmath}_{a}  D^I_{n-s}(g)=
-m\tr( gJ_0 g^{-1}J^{a})D^I_{n-s}(J_{a}g)=
m D^I_{n-s}(gJ_{0})= 2ms D^I_{n-s}(g).
\ee
The physical Hilbert space associated with the
particle is the usual Poincar\'e representation,
\be{\cal H}_{m,s} =\bigoplus_{I|I-s \in \N} V_{I}= \{D^I_{n-s}(g)|I-s \in
\N,\ |n|\leq I\}.
\ee
Instead of labelling the particle propagator by couples $\bar{g}_{i},\bar{g}_{f}$
we can label it by a pair of Lorentz indices $I_{i},I_{f}$.
This means that the particle propagator $G_{P_{m,s},I_{i},I_{f}}(\o,e)$ is
viewed as an operator in $\mathrm{Hom}(V_{I_i},V_{I_f})$.
The restriction we made on the value of $(\o,e)$ along the particle
world-line can be relaxed, it doesn't change our argument but just the
value  of $G_{P_{m,s},I_{i},I_{f}}$.
This shows,  in the case of particle insertions, that the propagator depends not only on a given
interpolating manifold $M$ but also on additional data characterizing the
evolution of the particles.
The labels $I_i,I_f$ are labels of the total angular momenta,
as we have seen, we can also understand them as labelling
the position of the particle in the direction transverse to the momenta
since $ I_f(I_f+1)-I_i(I_i+1)=m^2[ (x^2_\bot)_f-(x^2_\bot)_i]$.

\begin{figure}[ht]\psfrag{Gf}{$\G $}\psfrag{vb}{$\bar{v}$}
\psfrag{vt}{$\tilde{v}$}
\psfrag{v}{$v$} \psfrag{Dal}{$\da$}
\includegraphics[width=4cm]{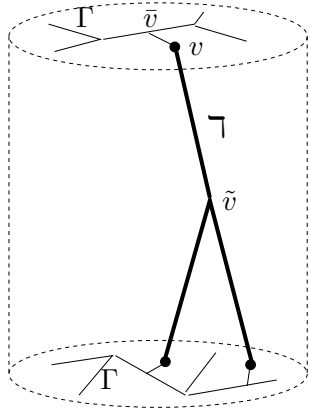}
\caption{boundary and particle graphs}\label{decorpartgraphs}
\end{figure}

We will consider the general case of interacting  particles of arbitrary spin.
In this case the data we need are encoded into what we will call a {\it decorated particle graph}
and will denote by $ \da_D $\footnote{ We could equivalently call these graphs {\it Feynman graphs}
since they label all possible Feynman graphs of 3d field theories, however
this terminology usually suppose that there is a specific field theory behind
the construction of this graphs and that we should sum over them.
It is an interesting problem to describe such a theory but this is not
the goal of this paper}.
$\da$ is graph embedded in $M$ such that its open ends $v$ are all lying in
$\partial M$.
$D=(m_e,S_e,I_{s_e},I_{t_e}, \imath_{\tilde{v}}) $ is a decoration of
$\da$ where
each edge $e$ of $\da$ is labelled by a mass and a spin $(m_e, S_e)$;
 each starting point of an edge $e$ is denoted by $s_e$ and labelled by a $\SU (2)$ representation
$V_{I_{s_e}}$, each terminal point of an edge $e$ is denoted by $t_e$ and labelled by a
$\SU (2)$ representation $V_{I_{t_e}}$; finally each internal vertex
$\tilde{v}$ is labelled by an intertwiner
$\imath_{\tilde{v}}$ contracting the Lorentz representations $I$
(see figure \ref{decorpartgraphs}).

Given a decorated manifold with boundary $M,\da_D$ we now want to describe
the construction of the kinematical spin network states and their transition
amplitudes.
% and identified  to the open ends of the spin network $\G,j_{\bar{e}},j_v)$.
We have to consider states that are not gauge
invariant at the location of the particle.
Such kinematical states are described by
spin networks with open ends \cite{BaezKrasnov}.
We denote by $\bar{v}$ the internal vertices of $\Gamma$ and
by $v$ the open ends of $\Gamma$,
they are identified  with the open ends of the particle graph
$\da$. We also denote by $\bar{e}$ the internal edges of $\Gamma$
and by $e_{v}$ the edges of $\Gamma$ ending on an open end $v$.
Given a graph $\Gamma$ whose internal edges are colored by $j_{\bar{e}}$ and
external edges colored by $I_v$, we can construct in the same spirit as before a spin network
functional $\F_{(\G,j_{\bar{e}},I_v)}(\bar{\o})$.
The difference with the previous case is that now
$\F_{(\G,j_{\bar{e}},I_v)}(\bar{\o})$ is not a scalar but take value in
$\otimes_v V_{I_v}$.
The construction is similar to the previous one, let us denote
by $ g_{\bar{e}}(\bar{\o})$ the holonomy along internal edges and
$ g_v(\bar{\o})$ the holonomy along external edges. The spin network functional
is given by
\be\label{Spinneto}
\F_{(\G,j_{\bar{e}},I_v)}(\bar{\o}) =
\bra \otimes_{\bar{v}} C_{\bar{v}} |\otimes_{\bar{e}} D_{j_{\bar{e}}}(g_{\bar{e}}(\bar{\o}))
\otimes_{v} D_{I_{v}}(g_{v}(\bar{\o}))\ket \in \otimes_v V_{I_v}.
\ee
The intertwiners $C_{\bar{v}}$ live only at internal vertices where a pairing occur, there is no contraction
at external vertices.
As before, we can consider the linear combination of all spin networks supported on a graph
and define the basis independent spin network Hilbert space
\be
{\cal{H}}_{\Gamma,I_v} = L^{2}\left((G^{|\bar{e}|+|v|}\rightarrow \otimes_v V_{I_v}) /G^{|\bar{v}|}\right),
\ee
where $|\bar{e}|$ denotes the number of internal edges of $\Gamma$,$|v|$ the number of open ends of $\Gamma$,
$|\bar{v}|$ the number of internal vertices and the action of $G^{|\bar{v}|}$ on $G^{|\bar{e}|}$
is given by the structure of the graph.
This space is our kinematical Hilbert space.

It should be clear, from the discussion of the classical particle, that the spin label
$I_v$ on open edges have the meaning of  the total {\it angular momenta} carried by the particle.
The individual vectors in the representation $V_{I_v}$ label the direction of the angular momenta.
$I_v$ {\it doesn't}  have the meaning of the spin of the particle, which is the component
of the angular momenta
along the direction of the particle momenta, this is important to stress since this is
often a source of confusion. The spin and the mass  appear dynamically
 in the choice of the propagator determining the physical scalar product.

Due to Lorentz gauge invariance not all choices of $I_v$ are admissible and when
a choice is admissible not all vectors in $\otimes_v V_{I_v}$ appear in the image
of $\F_{(\G,j_{\bar{e}},I_v)}$. The image of
$\F_{(\G,j_{\bar{e}},I_v)}$ lies in the invariant subspace $(\otimes_v V_{I_v})^G$ which is the
space of intertwiners $ \C \rightarrow \otimes_v V_{I_v}$, where $\C$ denotes the trivial representation of
$\SU (2)$.

The full propagator is not a scalar function but an operator depending on the decorated
 particle graph:
\be
G_{M,\da_{D}}(\bar{\o}_f,\bar{\o}_i):
\left(\otimes_{v_{i}} V_{I_{v_{i}}}\right)^G
\rightarrow \left(\otimes_{v_{f}} V_{I_{v_{f}}}\right)^G.
\ee
The amplitude is given by
\be\label{physprodspinpar}
\bra\F_{f}|\F_{i}\ket_{M,\da_D}
= \int \cD \bar{\o} \bra \F_f(\bar{\o}_f)|G_{M,\da_D}(\bar{\o}_f, \bar{\o}_i)|\F_i(\bar{\o}_i)\ket.
\ee
The measure $\cD \bar{\o}$ is the Ashtekar-Lewandowski diffeomorphism invariant measure \cite{Ashtekar:1994mh}.
Due to diffeomorphism invariance the propagator depends only on the diffeomorphism class of the embedding of
$\da$ in $M$.
In the case where $M= \Sigma \times I$, $\Sigma$ being a 2d surface with $n$ punctures,
 the physical scalar product on
${\cal{H}}_{\Gamma,I_v}$ is given by the following decorated graph:
$\da$ is the union of unbraided oriented segments $e_v$ joining $v_i$ with $v_f$.
The coloring of each line by $(m_{e_v},S_{e_v})$ labels the mass and spin of the particle,
also $I_{s(e_v)}=I_{v_i}$ and $I_{t(e_v)}= I_{v_f}$.

\section{The Ponzano-Regge model  and gauge fixing}\label{PR:model}
The Ponzano-Regge model is constructed from the continuum partition
function
\begin{equation}
Z_M=\int \cD \o \cD e \exp\le[\frac{i}{16\pi G}\int_\cM \tr\le(e\w F(\o)\ri)\ri],
\end{equation}
by considering a triangulation $\D$ (and its dual $\D^*$) of
$\cM$, and replacing the set of configurations variables by discrete
analogs in the spirit of lattice gauge theory. The connection field
is replaced by group elements $g_{e^*}$ associated to the dual
edges $e^*$ of $\D^*$, and representing the holonomy of the
connection field along these edges. The frame field is replaced by
Lie algebra elements $X_e$ associated to the edges $e$ of $\D$
and representing the integration of $e$ along these edges. The
curvature 2-form is now represented as group elements $G_e$ living
on the edges $e$ (or dual faces $f^*$), and obtained as the
ordered product of the group elements $g_{e^*}$ for dual edges
$e^*\subset f^*$, upon the choice of a starting dual vertex on the
dual face.

$\bullet$ The
discretized partition function is thus
\begin{equation} \label{eqn:discrete-PI}
Z_{PR}(\D)=\le(\prod_{e^*} \int_{\SU(2)}dg_{e^*}\ri)\le(\prod_e
\int_{\su(2)}dX_e\ri) \exp\le[i\sum_e \tr(X_e G_e) \ri].
\end{equation}
One can then integrate over the $X_e$ variables (see appendix (\ref{app:delta})),
\begin{equation}\label{eqn:discreteint-PI}
Z_{PR}(\D)=\le(\prod_{e^*} \int_{\SU(2)}dg_{e^*}\ri)\left(\prod_e \d(G_e)\right)
\end{equation}
where $\delta(g)$ is the delta function over the group\footnote{As it is explained
in the appendix (\ref{app:delta}) it is the delta function on $\SO (3)$ and not
$\SU(2)$. It is also explained how we can modify
(\ref{eqn:discrete-PI}) in order to get the delta function on $\SU(2)$.}.
The Ponzano-Regge partition function is recovered from
(\ref{eqn:discreteint-PI}) by expanding the $\d$ function using the
Plancherel decomposition, and then integrating over $g_{e^*}$
using recoupling identities for $\SU(2)$ \cite{spinfoam}.
The result is a realization of (\ref{eqn:continuum-PI}) and
  can be expressed as a summation over coloring of a product of 6j symbols.
\be \label{eqn:PRpart}
Z_{PR}(\D)= \sum_{\{j_e\}} \prod_e d_{j_e} \prod_t
\left\{
\begin{array}{ccc}
    j_{e_{t_{1}}} &  j_{e_{t_{2}}} &  j_{e_{t_{3}}} \\
    j_{e_{t_{4}}} &  j_{e_{t_{5}}} &  j_{e_{t_{6}}}
    \end{array}
    \right\},
    \ee
where the summation is over all edges of $\D$ and the product of 6j symbols is over all tetrahedra.
$d_j=2j+1$ denotes the dimension of the spin $j$ representation and
$e_{t_{i}}$ denotes the six edges belonging to the tetrahedra $t$.
$j_e$ plays the role of a duiscrete $X_e$, it is therefore interpreted as a length and the
Ponzano-Regge sum is a sum over all geometries supported by a triangulation.

$\bullet$ In the case of manifold with boundaries but no particles, we can define
the physical scalar product between spin network states $\F_{\Gamma_f,j_{\bar{e}_f}} \in {\cal H}_{\Gamma_f},
\F_{\Gamma_i,j_{\bar{e}_i}}\in {\cal H}_{\Gamma_i}$
in a similar way
\be
<\F_{\Gamma_f,j_{\bar{e}_f}}|\F_{\Gamma_i,j_{\bar{e}_i}}>_{PR}= \le(\prod_{e^*}
\int_{\SU(2)}dg_{e^*}\ri)\prod_e \d(G_e)\bar{\F}_{\Gamma_f,j_{\bar{e}_f}}(g_{\bar{e}_f})
\F_{\Gamma_i,j_{\bar{e}_i}}(g_{\bar{e}_i}),
\ee
the integral being over all group elements  associated with dual edges including the ones $\bar{e}_f,\bar{e}_i$ lying
 in the boundary. As before we can recover Ponzano-Regge like expressions
giving  the physical scalar product (\ref{physprod}) as a state sum model
\be \label{eqn:PRpartamp}
<\F_{\Gamma_f,j_{\bar{e}_f}}|\F_{\Gamma_i,j_{\bar{e}_i}}>_{PR}=
\sum_{\{j_e\}} \prod_e d_{j_e} \prod_t \left\{
\begin{array}{ccc}
    j_{e_{t_{1}}} &  j_{e_{t_{2}}} &  j_{e_{t_{3}}} \\
    j_{e_{t_{4}}} &  j_{e_{t_{5}}} &  j_{e_{t_{6}}}
    \end{array}
    \right\},
\ee
where the summation is over all edges of $\D$ which do not belong to the boundary, the coloring
of the boundary edges $\bar{e}_f, \bar{e}_i$ is fixed to be $j_{\bar{e}_f}, j_{\bar{e}_i}$.

$\bullet$ In the general case of manifold with boundaries and with particles we can also define a transition
amplitude which gives an explicit realisation of (\ref{physprodspinpar}). This one has never been written
before\footnote{some hints concerning the insertion
of spinless particles  were already given in \cite{ConfFK}} and the explicit description of it is one of
the main result of this paper. We will present its construction in detail in section \ref{tranpartamp}.
In the case of interacting spinless particles the result is very simple.
The particle graph $\da$ is living on the edges of the triangulation $\D$,
and we choose a decoration $D$ where all spins and angular momenta are $0$.
In this case the amplitude including boundaries and particles is given by
\be\label{spinlesspart1}
\bra \F_{\G_f,\j_f}|\F_{\G_i,j_i}\ket_{M,\da}=
\sum_{j_{e}}
\left(\prod_{e \notin  \da}d_{j_{e}} \right)
\left(\prod_{e \in \da} \chi_{j_{e}}(h_{\t_{e}})\right)
\prod_{t}
\left\{
\begin{array}{ccc}
    j_{e_{t_{1}}} &  j_{e_{t_{2}}} &  j_{e_{t_{3}}} \\
    j_{e_{t_{4}}} &  j_{e_{t_{5}}} &  j_{e_{t_{6}}}
    \end{array}
    \right\},
    \ee
 the summation is over all edges of $\D$ which do not belong to the boundary, the coloring
of the boundary edges $\bar{e}_f, \bar{e}_i$ is fixed to be $j_{\bar{e}_f}, j_{\bar{e}_i}$.
The edges in $\da$ carry a factor $\chi_{j}(h_{\t})$, which is the trace in the representation
of spin $j$ of the group element $h_{\t} = \exp( \t J_0)$ (\ref{ht}).
$\t$ is half the deficit angle created by the presence of a mass $m$,
$\theta = 4\pi G m $.

The general case involving interacting spinning particles is similar but more involved and
described in detail in section \ref{transpinpart}.
For all these amplitudes we can recover Ponzano-Regge like expressions
as state sum models.
However the naive definitions of these models  is plagued with divergences.
We need to do a careful analysis of the gauge symmetries of the discrete model
in order to understand these divergencies and to take care of them.
This is what we are presenting now.

\subsection{The gauge symmetries of the Ponzano-Regge model}
In this part we present the identification of the gauge symmetries
of the Ponzano-Regge model, and their gauge fixing. We discuss
also the case of a manifold with boundaries and insertion of
particles. This work completes the work begun in \cite{FLdiffeo}
where we only gauge fixed one of the symmetries. Note that we
corrected here the exact form of the action
(\ref{eqn:discrete-PI}) and as a consequence, the formula are
slightly different from \cite{FLdiffeo}.

\subsubsection{Discrete symmetries of the Ponzano-Regge model}
The gauge symmetries of the continuum action (Lorentz and
translation gauge symmetries) have been described in the first
section. The discrete analog of the Lorentz symmetry is parameterized
by group elements $k_{v^*}$ living at the dual vertices of the
triangulation. It acts as
\begin{eqnarray}
g_{e^*}&\to& k^{-1}_{t_{e^*}}g_{e^*}k_{s_{e^*}} \\
G_e&\to&k^{-1}_{st_e} G_e k_{st_e} \\ X_e&\to&k^{-1}_{st_e} X_e
k_{st_e}
\end{eqnarray}
where $s_{e^*}$ and $t_{e^*}$ denote the dual vertices source and
target of $e^*$, $st_e$ denotes the dual vertex which is the
starting point for computing the curvature on the dual face
$f^*\sim e$.

To express the discrete analog of the translational symmetry, we
observe that the discrete action appearing in
(\ref{eqn:discrete-PI}) can be written as
\begin{equation}
S=\tr(\sum_e X_e P_e)
\end{equation}
where $P_e= P_e^a {J}_a$ is the projection of $G_e$ on the Lie algebra and
is defined by
\begin{equation}
G_e=u_e Id+ P_e^a {J}_a,
\end{equation}
where $u_e^2 +P_e^a {P_e}_a =1$.
In the continuum we have seen that the translational symmetry is
due to the existence of Bianchi identity.
At the discrete level we also have a Bianchi identity for each vertex of
the triangulation:
\cite{Kawamoto:1999tf}
\begin{equation}\label{eqn:Bianchigroup}
\prod_{e\supset v} (k_v^e)^{-1} G_e^{\e (e,v)} k_v^e=Id,
\end{equation}
where the product is over all edges meeting at $v$, $k_v^e$ are group elements dependent
on $g_{e^*}$ characterizing
the parallel  transport from a fixed vertex to $st_e$ and
$\e(e,v)=\pm1$ depending on whether $e$ is pointing toward $v$ or in the opposite direction.
We can write this identity in terms of the $P_e$. One finds that there exist
Lie algebra elements $\O_e^v$ and scalars $U_e^v$ which can be expressed in terms
of the $P_e$'s for edges meeting at $v$, such that
\be
\label{eqn:discreteBianchi}
\sum_{e\supset v} \e(e,v) (k_v^e)^{-1}\le(U_e^v {P}_e+\co{\O_e^v}{P_e}\ri)k_v^e = 0.
\ee
It is easy to check that
the discrete analog of the translational symmetry is then
parameterized by Lie algebra elements $\F_v$ living at the vertices
of the triangulation and acts as
\begin{equation}\label{eqn:discretetrans}
X_e\to X_e+ \e(e,v) \le(U_e^v \left(k_v^e\F_v(k_v^e)^{-1}\right)-
[\O_e^v,\left(k_v^e\F_v(k_v^e)^{-1}\right)]\ri)
\end{equation}
for $v\subset e$.

\subsubsection{Gauge fixing of the symmetries}
Having identify the discrete gauge symmetries of the discrete
action (\ref{eqn:discrete-PI}), it has been observed
\cite{FLdiffeo} that the infinite gauge volume of the
translational symmetry was actually responsible for the
divergencies of the Ponzano-Regge model. We then proposed a
procedure to gauge fix this symmetry. The Lorentz gauge group
being compact in euclidian case, we ignored the gauge fixing of
the Lorentz symmetry and concentrated on the translational
symmetry. Having in mind the possible applications for the
Lorentzian case where the Lorentz group is no longer compact, we
now need to be able to perform both gauge fixing at the same time
in a consistent way. This is the purpose of this part.

Both  discrete gauge symmetries act at the vertices of a
graph (1-skeleton and dual 1-skeleton of $\D$), on variables
living on the edges of this graph. The usual method to gauge fix
such a symmetry is to choose a maximal tree, i.e. a connected
subgraph touching every vertex without forming a loop, then to use
the gauge symmetries at the vertices to gauge fix all the
variables living on the edges of the tree. This method has been
applied to the Lorentz symmetry as a gauge fixing for the
non-compact spin-networks \cite{FLnoncomp}, and to the
translational symmetry for the Ponzano-Regge model
\cite{FLdiffeo}. Recall that this procedure uses the action of the
gauge symmetries at every vertex except one, taken as the root of
the tree. This gives rise to a remaining global symmetry that has to
be studied in a second time.

$\bullet$ We want to perform both gauge fixing. We thus choose a maximal
tree $T$ in the 1-skeleton of the triangulation $\D$ and a maximal
tree $T^*$ in its dual 1-skeleton. If $|v|$ and $|t|$ are the number
of vertices in the 1-skeleton and dual 1-skeleton, these trees
contain respectively $|v|-1$ and $|t|-1$ edges. We use the gauge
symmetries to fix
\begin{eqnarray}
g_{e^*}=1,\ \ && \forall e^*\in T^*, \\ X_e=0,\ \ && \forall e\in T.
\end{eqnarray}

We now need to identify the corresponding Fadeev-Popov
determinant. The jacobian for fixing the translational symmetry
has been computed in \cite{FLdiffeo}. Using the fact that $T^*$ is
a tree, one can show that the jacobian associated to the gauge
fixing of the Lorentz symmetry is 1. Finally we have for the
Fadeev-Popov determinant
\begin{equation}
\D_{FP}=\prod_{e\in T} ((U^{t_{e}}_{e})^{2}¥+|\O_e^{t_e}|^2)\le|U_e^{t_{e}}\ri|.
\end{equation}
Using the argument given in \cite{FLdiffeo} we can prove that this
determinant is actually 1 while included in the partition
function. This is due to the fact that the product of $\d(G_e)$ for
$e\not\in T$ is enough to impose $P_e=0$ on every edge as long as
$T$ does not touch the boundary or the particle. The
net result is that there is no Fadeev-Popov appearing in the
partition function after the double gauge fixing.
After this gauge fixing procedure we have
\begin{equation}\label{gfparf}
Z[\D\backslash\{T,T^*\}]=\le(\prod_{e^*\not\in T^*}
\int_{\SU(2)}dg_{e^*}\ri)\prod_{e\not\in T} \d(G_e).
\end{equation}

$\bullet$
After this gauge fixing procedure, we are left with a
triangulation which possesses only one vertex and one dual vertex.
However, there are still some invariances associated. The remaining
Lorentz invariance acting at the unique dual vertex is a diagonal
$AdG$ invariance $g_{e^*}\to k^{-1}g_{e^*}k$. It has been shown in
\cite{FLnoncomp} that this remaining invariance can be gauge fixed
using the measure
\begin{equation}
d\u(g_1,...,g_N)=d\u(g_1,g_2)dg_3...dg_N,
\end{equation} where $dg$ is the Haar measure on $G$ and
$d\u(g_1,g_2)$ is defined by
\begin{equation}\label{eqn:measureFL}
\int_{G^2} d\u(g_1,g_2) f(g_1,g_2) = \int_{H \times G/H} dh dx,
f(h,s(x))
\end{equation}
where $H$ is the Cartan subgroup of $\SU(2)$ and $s:G/H \rightarrow G$ is a
given section\footnote{We present
here only the measure in the case of a unique Cartan subgroup,
relevant for $\SU(2)$. The case of many Cartan subgroup relevant
for $\SL(2,\R)$ can be treated in the same way, see
\cite{FLnoncomp}}. This choice of measure allows to gauge fix the
remaining Lorentz invariance.

\vspace{1ex}

Let us now consider the case of the remaining translational
symmetry. This symmetry is supposed to arise from the Bianchi
identity around the last vertex. As this vertex is the last one,
every edge starts and ends on it. The corresponding Bianchi
identity (\ref{eqn:Bianchigroup}) thus involves both $G_e$ and
$G_e^{-1}$. The key point is to understand the order of the
elements in (\ref{eqn:Bianchigroup}). This order is related to the
topology of the neighborhood of the vertex. Since we started from
a triangulation of a manifold, the neighborhood of each original vertex
was a three-sphere. Now we
remove the edges corresponding to a tree, which by definition has
no loops. Hence its tubular neighborhood has the topology of a
sphere and we are left with a last vertex whose neighborhood is
still a 3-sphere. The Bianchi identity is trivial which means that
there is no residual action of the translational symmetry. This
just means that our original parameterization of the gauge group
was redundant, and that the gauge group was actually ${\su (2)}^{|v|-1}$.

\subsubsection{Gauge fixing: a mathematical argument}

There is a beautiful and simple mathematical argument which allows
to understand the necessity of gauge fixing which goes as follows.
We know that in the theory of integration what we should integrate
over a manifold of dimension $n$ is not a function but a density
of weight $n$. A delta function on the group $\SU(2)$ is  a
distribution, that is a density of weight $3$, i-e the only thing
that make sense is the quantity $dg \delta(g)$ and not
$\delta(g)$. Before gauge fixing, the integration manifold
 is a manifold of dimension $|f|$, the
number of faces of $\Delta$ and the integrand is a density of
weight $|e|$, the number of edges of $\Delta$. The dimension of the
space does not match the density weight of the integrant and the
integration does not really make sense. After
gauge fixing the density weight of the integrant (\ref{gfparf}) is $3(|e| -|v| +
1)$ and the dimension of the integration manifold is $3(|f| -|t| + 1)$, where
$|v|$ (resp. $|t|$) denotes the number of vertices (resp.
tetrahedra) of $\Delta$. The difference between the two is given
by $$-3\chi = 3(|t|-|f|+|e|-|v|),$$ where $\chi$ is the Euler
characteristic of the 3d manifold $M$. If $M$ is a closed manifold
its Euler characteristic is zero, therefore the integral
(\ref{gfparf}) is a priori well defined. If $M$ is not
closed $\chi(M) = 1/2\chi(\partial M) =(1-g)$, where $g$ is the
genus of the boundary if $\partial M$ is connected. Therefore if
the boundary does not contain any sphere or torus $-3\chi(M) =
3(g-1)$ is strictly positive. The result of integration in this
case is a density of positive weight $3(g-1)$ which is exactly
half the dimension of the moduli space of flat $\SU(2)$ connection
on $\partial M$. This is the correct result since we will see that
in the case of manifold with boundaries, the quantity
(\ref{gfparf}) should be interpreted as the integration
kernel for quantum transition amplitudes. One should be careful
however that the counting argument we have just given is naive and
does not always guarantee the definiteness of the
integral (\ref{gfparf}) \footnote{In a similar way as having a negative naive
superficial divergences degree of a Feynman integral does not
always guarantee its convergence.}.

In terms of the mathematical argument given at the beginning of
this part, the result of this integration can be evaluated to be
\be
\label{mathgf}
\left|\frac{\wedge_{e^{*}\not\in T^{*}}
dg_{e^*}}{\wedge_{e\not\in T} dG_{e}}\right|_{G_{e}=0}, \ee where
$\wedge$ denotes the wedge product of forms. When the manifold is
closed this is just a jacobian, when the manifold admits boundaries
this is a form of positive degree by the previous argument. This
argument does not guarantee the definiteness of this jacobian since
there still could be situations in which $(g_{e^*})_{e^{*}\in
\Delta^*}$ does  not depend on some component of $(g_{e})_{e\in
\Delta}$. In this case the jacobian will be infinite.{ We will
analyze more precisely the finiteness issues in section \ref{closem}}.

\subsubsection{Inclusion of boundaries and particles}

We now discuss the generalization of the gauge-fixing procedure to
the case of a manifold with boundaries, when we try to compute
transition amplitudes between spin-networks states. We have seen
that at the classical level the Lorentz symmetry also acts on the
boundary, while the translational symmetry is such that the
parameter of transformation has to vanish on the boundary. This
is translated into the fact that the physical scalar product
between spin-network states
\begin{equation}
<\F_{\Gamma_f,j_{\bar{e}_f}}|\F_{\Gamma_i,j_{\bar{e}_i}}>_{PR}=
\le(\prod_{e^*} \int_{\SU(2)}dg_{e^*}\ri)\ \prod_e \d(G_e)\
\bar{\F}_{\Gamma_f,j_{\bar{e}_f}}(g_{\bar{e}_f})
\F_{\Gamma_i,j_{\bar{e}_i}}(g_{\bar{e}_i}),
\end{equation}
has a Lorentz symmetry acting at all the dual vertices of the
triangulation (including the dual vertices lying on the boundary)
while the translational symmetry is restricted to the bulk
vertices. The gauge fixing of such amplitudes involves then a
maximal tree $\cT$ of the triangulation, touching at most one
vertex on the boundary (recall that there is one vertex in the
maximal tree that we actually don't use for the gauge fixing). The
gauge fixing of the Lorentz symmetry involves a dual maximal tree
$\cT^*$ touching every dual vertex, including those on the
boundary.

As discussed in the classical case, it is expected that the
introduction of a particle will break a part of the gauge
symmetries of gravity, turning them into particle degrees of freedom
 and modify the discussion of the gauge
fixing. Actually, the inclusion of a particle on an edge $e$ breaks
the translational gauge symmetry as does the inclusion of mass
classically. As we will discuss explicitly later, the discrete
transformation (\ref{eqn:discretetrans}) is not a symmetry of the
discrete action anymore.
A maximal tree $\cT$ for the gauge fixing
of the translational symmetry has to be chosen outside the
vertices lying on the trajectory of the particle.

\subsection{BRST invariance}
We have computed the gauge fixed partition function in terms of a gauge fixing procedure
involving two maximal trees. In a general gauge theory the gauge symmetry is obviously no longer visible
after gauge fixing. However there is an in-print of this symmetry in the gauge fixed theory
which is BRST invariance  assuring us that the gauge fixed partition function is independent
of the gauge choice.
In our case we have to check if the gauge fixed partition function
\begin{equation}\label{PRpartfun}
Z[\D\backslash\{T,T^*\}]=\le(\prod_{e^*\not\in T^*}
\int_{\SU(2)}dg_{e^*}\ri)\prod_{e\not\in T} \d(G_e),
\end{equation}
depends on the choice of $(T,T^*)$. It is shown in a companion
paper \cite{mathDL} that the gauge fixing procedure is independent
of the choice of maximal trees i.e.
\begin{equation}
Z[\D\backslash\{T_1,T_1^*\}]=Z[\D\backslash\{T_2,T_2^*\}].
\end{equation}
This property is a remnant of the original gauge + translational
symmetry. It follows from the invariance properties of the
path-integral but is proved more easily using the graphical tools
introduced in \cite{mathDL}.
It is also proved in this paper that the partition function
does not depend on the choice of the triangulation matching the boundary data
so it is a topological invariant
\begin{equation}
Z[\D_1\backslash\{T_1,T_1^*\}]=Z[\D_2\backslash\{T_2,T_2^*\}].
\end{equation}

\section{Ponzano-Regge Observables}\label{Oanddual}

In this section we  introduce the notion
of Ponzano-Regge observables that are discrete analog
of `kinematical observables' presented in section \ref{transamp}.
We show that the gauge fixing procedure can be written
as the evaluation of the expectation value of a gauge fixing
operator. We also present the Ponzano-Regge observables realizing the insertion of particles.
We describe the link between Ponzano-Regge model and Chern-Simons theory
and show a remarkable duality property between gauge fixing and insertion of particles.

First, let us remark that if one uses the Plancherel
identity \be \delta(g) =\sum_j\frac{d_j}{V_G} \chi_j(g), \ee where $j$ is a spin,
$\chi_j(g) =\tr_{V_j}(D_j(g))$ is
the character of the spin $j$ representation, $d_j$ is the dimension of the representation
 and $V_G$ is the volume of the group, then
 the partition function (\ref{PRpartfun})
associated with a triangulation $\D$ can be written as
\be
\sum_{j_e} \frac{d_{j_e}}{V_G} \int \prod_{e^*}dg_{e^*} Z_\D(j_e,g_{e^*}),
\ee
 where,
\be
Z_{PR}(\D)=Z_\D(j_e,g_{e^*})= {\prod}_{e^*\notin T^*}
\chi_{j_e}(G_e),\, G_e=\overleftarrow{\prod}_{e^*\subset e}g_{e^*}^{\e(e,e^*)},
\ee
where ${\e(e,e^*)}=\pm 1$ is the relative orientation of $e^*$ with respect to $e$.
 We will call PR-observable \footnote{PR stands for Ponzano-Regge of course, these observables
 are not in general gauge invariant and in this case they are `kinematical observables' }
 a function $\cO(j_e,g_{e^*})$
and define its expectation value as
\begin{equation}
\bra\cO\ket_\Delta=\sum_{j_e} \frac{d_{j_e}}{V_G} {\prod}_{e^*\in T^*}\int dg_{e^*} Z_\Delta(j_e,g_{e^*})
\cO(j_e,g_{e^*}).
\end{equation}

\subsection{ Gauge fixing observables}
In particular, an important class of PR-observables are those fixing
the lengths along a graph $\G$ of $\D$
\begin{equation}
\cO_{\G,j^\G_e}(j_e,g_{e^*})=\prod_{e\in\G} \frac{\d_{j_e,j_e^\G}}{d_{j_e^\G}},
\end{equation}
where $\delta_{j,j'}$ is the Kronecker delta function.
And those fixing the holonomy along a dual graph $\G^*$
\begin{equation}
\cO_{\G^*,\t^{\G^*}_{e^*}}(j_e,g_{e^*})=\prod_{e^* \in\G^*}
{\d_{\t^{\G^*}_{e^*}}(g_{e^*})},
\end{equation}
where $\t\in [0,2\pi]$ and $\d_{\t}(g)$ is a delta function
fixing $g$ to be in the same conjugacy class as
$h_\t= \exp(2\t J_0)$\footnote{the normalization and other properties of this function are described
in the appendix \ref{app:su2} }.
More generally, we can define operators fixing edges length along a
graph $\G$ of $\D$ and the conjugacy class of $g_e^*$ along a dual
graph $\G^*$ of $\D^*$.
\be
\cO_{(\G,j^\G_e)(\G^*,\t^{\G^*}_{e^*})}(j_e,g_{e^*})= \prod_{e\in\G}
\frac{\d_{j_e,j_e^\G}}{d_{j_e^\G}} \prod_{e^* \in\G^*}
{\d_{\t^{\G^*}_{e^*}}(g_{e^*})}.
\ee
In general the evaluation of the expectation value of
$\cO_{(\G,j^\G_e)(\G^*,\t^{\G^*}_{e^*})}$ are quite involved;
however if $\Gamma=T$ and $ \Gamma^*=T^*$ are trees in $\D$ and $\D^*$ then
the evaluation simplifies drastically, we have the following
\cite{mathDL}
\be \label{gaexp}
\bra \cO_{(T,j^T_e)(T^*,\t^{T^*}_{e^*})}\ket_\Delta =
\left(\prod_{e \in T}d_{j_{e}^T}\right)
\left(\prod_{e^*\in T^*} \frac{V_G}{V_H}\right)\
Z[\D\backslash\{T,T^*\}], \ee
where  $V_G, V_H$ denote the volume of
G and its Cartan subgroup $H$.
This shows that the degrees of
freedom living along trees or dual trees decouple from the
bulk degrees of freedom. We have seen in the precedent section that
indeed, such degrees of freedom are pure gauge and therefore the
expectation value of PR-observables depending only on such degrees of
freedom is expected to factorize. Equation (\ref{gaexp}) show that
such expectation value does not depend for instance on the topology
of the manifold $M$ triangulated by $\Delta$ or the decoration on
the boundary of $M$. Such PR-observables are therefore not physical
observables, they couple only to the gauge degrees of freedom and not
to the physical degrees of freedom.
If we consider the observable $\cO_{\G,j^\G_e}$ where $\Gamma$ is not a tree
then this PR-observable will couple to the dynamical degree of freedom
of the theory and it is a physical observable giving us details about
the geometry of our space-time.
Due to gauge invariance two PR-observables $\cO_{\G_1,j^{\G_1}_e}$
$\cO_{\G_2,j^{\G_2}_e}$ will carry the same physical information
if $(\G_1,j^{\G_1}_e)$ and $(\G_2,j^{\G_2}_e)$ are diffeomorphic colored graphs
or if they can be collapsed using a tree to diffeomorphic colored
graphs.

An other interesting evaluation of the expectation value of this
operator is when $\G$ is the one-skeleton of $\D$ (in this case we
denote it $\D$, by an abuse of notation) and when $\G^{*}¥=\D^{*}¥$ is
the one skeleton of $\D^{*}¥$.
In this case we introduce a special  notation for this object
\be\label{Zlink}
Z_{\D}(j_{e},\t_{e^{*}¥})= \bra \cO_{(\D,j_e)(\D^*,\t_{e^*})}\ket_\D.
\ee
This object does not contain anymore any summation over the length of
the edges and integration over the conjugacy class of $g_{e^{*}¥}$,
its evaluation is therefore trivially finite since
\be
Z_{\D}(j_{e},\t_{e^{*}¥})= \int_{G/H}\prod_{e^*\in \Delta^*}dx_{e^*}
\prod_{e\in\Delta} \chi_{j_e}(\overleftarrow{\prod}_{e^*\subset
e}x_{e^*}h_{\theta_{e^*}}^{\e(e,e^*)}x_{e^*}^{-1})
\ee
where $dx$ is the induced measure on $G/H$ and $h_{\theta}$ is defined in (\ref{ht}).
This formula follows from the Weyl integration formula (see appendix \ref{app:su2}).

\subsection{Chern-Simons formulation}
 The remarkable property of this object, which is proven in detail in
\cite{mathDL}, is that $Z_{\D}(j_{e},\t_{e^{*}¥})$
can be interpreted as a quantum group evaluation of a chain mail link $L_{\Delta}$.
The link in question $L_{\Delta}$ is described in detail in \cite{mathDL}
but can be sketchily described as follows:
One can consider the handlebodies $H$ and $H^*$
respectively built as the thickening of the 1-skeleton and dual
1-skeleton of $\D$. The gluing of  $H$ and $H^*$ along their boundaries
gives a Heegard splitting of the manifold $\cM$ \cite{robertsSKTV}.
One can now consider the meridian circles of $H$ and $H^*$,
and by slightly pushing the meridians of $H$ into $H^*$,
one obtains the link $L_\D$. It should be noted that this
link is made from two different kinds of components: the ones coming
from $H$ and the others coming from $H^{*}$.
The quantum group in question here is a kappa deformation of
the Euclidean Poincar\'e group $ISO(3)$, it is denoted $D_{\kappa}(\SU(2))$
\footnote{$\kappa= 1/4G$ is the Planck mass in three dimensional gravity}
and can
constructed as a Drinfeld double of $\SU(2)$ \cite{baismuller}.
Its relevance in the context of 2+1 gravity as been recently analyzed
in \cite{FLeeJerzy}.
The  link is colored by spins $j$ for the components
coming from $H$ and angles $\theta$ for the $H^{*}$ components.
$j$ label pure spin representations of the kappa Poincar\'e group whereas
$\theta$ is interpreted as the mass of a spin zero representation.

It is well known that, at the classical level, three dimensional
gravity can be expressed as a Chern-Simons theory where the gauge group
is the Poincar\'e group.
The Chern-Simons connection $A$ can be written in terms of the spin
connection $\o$ and the frame field $e$, $A = \o_{i}J^i+ e_{i}P^i$
where $J^i$  are rotation generators and $P^i$ translations.¥
Moreover, since the work of Witten \cite{WittenJones}, it is also
well known that quantum group evaluation of colored link gives a computation
of expectation value of Wilson loops in Chern-Simons theory.
Our result therefore gives an exact relation, {\it at the quantum level}, between
expectation value in the Ponzano-Regge version of three dimensional gravity and
the Chern-Simons formulation.
More precisely, given a link $L_{\D}$ whose components are colored by
$j_{e},\theta_{e^{*}}$ we can construct a Wilson loop functional
$W_{L_{\D},j_{e},\theta_{e^{*}}}(A)$ by taking the ordered exponential
of the Chern-Simons connection along the components of $L_{\D}$ and tracing
the result in the Poincar\'e representations labelled by $j_{e}, \theta_{e^{*}}$.
If we denote by $\bra \cdot \ket_{CS}$ the expectation value in the
Poincar\'e Chern-Simons theory the result proved in \cite{mathDL}
reads
\be
\bra \cO_{(\D,j_e)(\D^*,\t_{e^*})}\ket_\D= Z_{\D}(j_{e},\t_{e^{*}¥})=
\bra W_{L_{\D},j_{e},\theta_{e^{*}}}(A) \ket_{CS}.
\ee
Note that the full partition function can be recovered as a sum
\be
Z_{PR}(\D)=
\sum_{j_e}\frac{d_{j_{e}}}{V_G} \int_{H/W} \prod_{e^{*}}d\t_e^*
\Delta^2(\theta_{e^{*}}) Z_{\D}(j_{e},\t_{e^{*}¥}).
\ee
From the Chern-Simons point of view this gives a surgery
presentation of the partition function.

\subsection{Particle observables}
Let us define another type of observables
\be\label{tildeO}
\tO_{\da,\t_e}(j_e,g_{e^{*}})=\prod_{e\in\da}
{\d_{j_e,0} \d_{\t_e}(G_e)},
\ee
where $\t_{e} \in [0,2\pi]$.
The expectation value of this PR-observable
removes (with $\d_{j_e,0}$) the flatness condition around the
edges of $\da$ imposed by the partition function, and fixes of
the holonomies  around the edges of $\da$ to be in the conjugacy class
of $h_{\t_{e}}$ (\ref{dphi}).
We will justify in greater detail in section \ref{tranpartamp} that the insertion
of this operator can be interpreted as the
presence of a spinless\footnote{more precisely this correspond to a particle which do not carry any
angular momenta} particle of mass $m_e = \t_{e}/4\pi G $
on the edge $e$. For instance, one can prove that the mass is conserved at bivalent vertices of $\da$.
One can also check that the expectation value of (\ref{tildeO}) is given by the expression
(\ref{spinlesspart1}).

Spinning particles can also be introduced
by the insertion of a PR observable described precisely in section \ref{transpinpart},
The structure of the observable is as follows.
First we consider
 a decorated particle graph $\da_D$ (see section \ref{spinningpart}) living along the edges of
 the triangulation $\Delta$, we introduce the
operator defined for each edge of the particle graph
\be
\tO_{\t_e,S_e,I_{t_e},I_{s_e}}(j_e,g_{e^{*}})= \delta_{j_e,0} \int_{G/H}du_e \delta(G_e u_e h_{\t_e}u_e^{-1})
D^{I_{t_e}}(g^*_{t_e}u_e)P^{S_e}_{I_{t_e},I_{s_e}}D^{I_{t_e}}\left((g^*_{s_e}u_e)^{-1}\right)
\ee
where $\t_e$ is the deficit angle associated with the mass $m_e = \t_e /8\pi G$, $S_e$ is the spin,
$s_e,t_e$ are the source and target of the edge $e$, $I_{s_e},I_{t_e}$ are Lorentz indices,
$D^{I}(g)$ is the representation matrix of the group element $g$ in the representation of spin $I$,
$P^{S_e}_{I_{t_e},I_{s_e}}=|I_{t_e},S_e><I_{s_e},S_e|$ where
$|I,S>$ is the vector of magnetic moment $S$ in the representation $I$
and $g^*_{t_e}$ is the group element associated with a path  going from $st(e)$ to $t_e^*$ (see section
\ref{spinningpart}).
Associated with each internal  vertex of $\da_D$ we have an intertwiner $i_{\tilde{v}}$ that can be used to
contract the open indices of $\tO_{\t_e,S_e,I_{t_e},I_{s_e},}$ to get
\be
\tO_{\da_D}=\prod_{e\in\da} \bra \otimes_{\tilde{v}} i_{\tilde{v}}|\otimes_e \tO_{\t_e,S_e,I_{t_e},I_{s_e},}\ket.
\ee
The expectation of this PR-observable
removes (with $\d_{j_e,0}$) the flatness condition  imposed by the partition function around the
edges of $\G$,  fixes of
the holonomies  around the edges of $\G$ to be in the conjugacy class
of $h_{\t_{e}}$ (\ref{dphi}) and insert a projector forcing the particle to be
of spin $S_e$.
On top of a mass conservation rule there is also a spin conservation rule.
In the case were all $I$'s and $S$'s are zero we recover the operator
(\ref{tildeO}).

\subsection{Wilson lines and time observable}
Before presenting a detailed justification of the introduction
of these massive spinning  observables corresponding to the inclusion
of massive spinning interacting particles it is of interest to
describe  other PR-observables.

The first one we would like to consider is what we call the identity
PR-observable. Let $\Gamma$ be a graph living in $\D$
we denote
\be
I_\G = \prod_{e\in \G} \delta_{j_e,0} \delta(G_e).
\ee
Such an observable removes (with $\delta_{j_e,0}$) the flatness condition
around the edges and insert back this condition (with $\delta(G_e)$).
It clearly does not change anything and does not depend on $\G$.
It is therefore neither  physical nor observable, its only interest lies
in the fact that it appears naturally when we consider the particle observables
associated with a zero mass and zero spin  propagating along $\G$, it
is therefore not a surprise that it is trivial.

 More interesting PR-observables are the Wilson observables. Given a
 graph $\Gamma^*\in \D^*$ whose edges are colored by spins $j_{e^*}$, we can define
 as in section \ref{Quamp} the spin network functional $\Phi_{(\Gamma^*,j_{e^*}^{\G^*})}$
which depends only on group elements $g_{e^*}$, $e^*\in \Gamma^*$.
The Wilson PR-observable is given by
 \be\label{wilsonline}
 W_{\Gamma^*,j_{e^*}^{\G^*}}(j_e,g_{e^*})=
 \prod_{e\in \Gamma^*}
\Phi_{(\Gamma^*,j_{e^*}^{\G^*})}(g_e^*)
\ee
Since this observable depends both on the spin connection
and the spin on the graph it is sensitive to the curvature and metric degrees of freedom.

The last observable we want to describe here, is by far the most interesting since
it allows to recover the notion of time at the quantum level.
One should however say that since we are working in the Euclidean context
there is no notion of time orientability and we cannot distinguish $T$ and $-T$,
this will be different in the Lorentzian context which will be the subject of another
work \cite{DLLor}.
%So, when we talk about time here we really mean modulo time orientability.

First, we can define the time operator only in the case where the manifold $M$
possesses boundaries $\partial M = \Sigma_i \coprod \Sigma_f$.
In this case we choose $L_n \in \D$ to be a succession of edges of $\D$;
 $L=(e_1,e_2,\cdots e_n)$ such that $s(e_{i+1}) = t(e_i)$ and $s(e_1)=v_i \in \Sigma_i$,
$t(e_n)=v_f \in \Sigma_f$. We also suppose that $L_n$ is not knotted, it is a simple line going from
$v_i$ to $v_f$ which are two vertices of the boundary triangulation.
We consider the PR-operator
\be\label{Time}
{T}_{L_n}(j) = \delta_{j_{e_1},j} \prod_{i=2}^n \delta_{j_{e_i,0}}.
\ee
One can check that the expectation value of this operator
does not depend on the number of edges composing $L_n$ or which edge we choose
to be of length $j$.
Ideally we would like to define the proper time as the distance
between the slices $\S_i$ and $\S_f$ along a minimal curve
with a fixed  initial direction.
The insertion of the operator $T(j)$ in the amplitude amounts to fix the
distance between the two slices but ignoring the information about
the direction of the curve.
It computes a superposition over all initial directions of  amplitudes for which the distance
between the two slices is $j$. This will be clear in our hamiltonian analysis.

\subsection{Duality properties}
In this section we would like to present a remarkable duality property between the
particle PR observables and the
gauge fixing PR observables.
\begin{prop}
Consider a graph $\G$ in $\D$. The observables $\cO_{\G,j^\G_e}$
and $\tO_{\G,\ff_e}$ are dual by Fourier transform in the following
sense
\begin{eqnarray}
<\cO_{\G,j^\G_e}>&=&
\int \prod_{e\in\G} d\ff_e \sin\le[(2j_e^\G+1)\ff_e\ri] \
\D({\ff_e}) <\tO_{\G,\ff_{e}}>,
\label{eqn:duality1}
\\ \D({\ff_e}) <\tO_{\G,\ff_e}>&=&\frac{2}{\pi}\sum_{\{j_e^\G\}} \prod_{e\in \G}
\sin\le[(2j^\G_e+1)\ff_e\ri] <\cO_{\G,j^\G_e}>,
\label{eqn:duality2}
\end{eqnarray}
where $\D(\t)=\sin \t$.
\end{prop}

\proof Let us prove the first formula. The RHS is written
\begin{equation}
RHS=\sum_{\{j_e,e\notin\G\}} \int \prod_{e^{*}}dg_{e^*}
\left(\int \prod_{e\in\G} d\ff_e \sin\le[(2j_e^{\G}+1)\ff_e\ri]
\frac{\D(\ff_e)}{V_{G}} {\d_{\ff_e}(G_e)}\right)
\prod_{e\notin\G}
\frac{d_{j_e}}{V_{G}}\X_{j_e}(G_e),
\end{equation}
where $G_e=\prod_{e^*\supset e}g_{e^*}$. In the following
it is always understood that $e\in \D, e^*\in\D^{*}$ unless otherwise
specified.
We first write
\begin{equation}
\sin \le[(2j_e^{\G}+1)\ff_e\ri]=
\frac{\sin \le[(2j_e^{\G}+1)\ff_e\ri]}{\sin\ff_e} \times \sin\ff_e
=\X_{j_e^{\G}}(h_{\ff_e}) \times \D({\ff_e}),
\end{equation}
and then integrate $\d_{\ff_e}(g_e)$ over $\ff_e$
 using the identity (\ref{deltaHW}).
One obtains
\begin{equation}
RHS=\sum_{j_e,e\notin\G} \int \prod_{e^{*}}dg_{e^*}
\left(\prod_{e\in\G} \frac{\X_{j_e^{\G}}(G_e) }{V_{G}}\right)
\left(\prod_{e\notin \G} \frac{d_{j_e}}{V_{G}}\X_{j_e}(G_e)\right),
\end{equation}
which is
\begin{eqnarray}
    RHS & =& \sum_{j_e} \int \prod_{e^{*}} dg_{e^*}
 \left(\prod_{e\in\G} \frac{\d_{j_e^{\G}}(j_e)}{d_{j_e^\G}}\right)
 \prod_e \frac{d_{j_e}}{V_{G}}\X_{j_e}(G_e)\\
&=&<\cO_{\G,j_e^\G}>=LHS,
\end{eqnarray}
and prove the first duality formula. The other one follows from
the first one and the orthogonality relation
\begin{equation} \sum_{d_j=1}^{\infty}
\sin(d_j\t)\sin(d_j\f)=\frac{\pi}{2}(\d(\t-\f)-\d(\t+\f)).
\end{equation}

\noindent\textit{Example :} A simple application of this duality
formula is given by the case where $M=S^3$, $\D$ being its  simplest
triangulation with two tetrahedra, one tetrahedra being the
interior of a 3-ball in $S^3$ the other one the exterior.
We take  $\G$ to be the 1-skeleton of the
triangulation, this is a tetrahedral graph.
In that case, the gauge fixing PR-observable is  the
square of the $6j$-symbol,
\be
<\cO_{\G,j_e}>_{\D}=\left\{
\begin{array}{ccc}
    j_{e_{{1}}} &  j_{e_{{2}}} &  j_{e_{{3}}} \\
    j_{e_{{4}}} &  j_{e_{{5}}} &  j_{e_{{6}}}
    \end{array}
    \right\}^2,
\ee
while the particle PR-observable is related to the Gram determinant
\be
<\tO_{\G,\ff_e}>_{\D} =\frac{1}{\sqrt{\det
(\cos\ff_e)}}.
\ee
This is the measure for $G^4/(G\times AdG)$, the duality
between these observables was first presented in \cite{FL6j}.
The above proposition is a generalization of the structure which was
discovered there. It is also a generalization to the classical
case of the duality relation proposed by Barrett \cite{Barrettduality}.

\section{Hamiltonian formulation and time evolution}\label{Hamfor}
In this section we present a hamiltonian formulation
of three dimensional gravity which generalizes the formalism
of Waelbroeck \cite{Waelbroeck:dt} and allows the inclusion of
spinning particles (more details will be given in \cite{FZapa}).
 We also briefly discuss its quantization and
give an Hamiltonian interpretation to the
PR amplitude with time.

\subsection{Classical analysis}
We consider a 2dimensional cauchy surface with punctures and denote
$\Delta$ one of its  triangulation.
The graph dual to $\D$ is an oriented trivalent graph with open ends
ending at the punctures
We can
construct the phase space of 3d gravity by assigning to each oriented edge
$e$  of $\Gamma$ a set of variables $(X_{e},g_{e})$.
We  call $\bar{e}$ the internal edges and
$e_{v}$ the edges starting from an open end $v$.
$g_{e}\in\SU (2)$ denotes
the parallel transport of the spin connection along the edge and
$X_{e}= X_{e}^a J_{a}$ is a Lie algebra element\footnote{$J^a=J_{a}=i\sigma$,
where $\sigma_a$ are the Pauli matrices, $\sigma_a \sigma_b = \delta_{ab}
+i \e_{abc} \sigma_{c}$, is
a basis of the Lie algebra}
characterizing the direction and length of the edge
 of the triangulation dual to $e$.
More precisely, we can obtain $X_{e}$ by
the integration of the frame field along edge of the triangulation
dual to $\G$. $X_{e}^a=\int_{{e}^*} dx^\mu g_{s_{e}}^x
e_{\mu}^a (g_{s_{e}}^x)^{-1}$ where $g_{s_{e}}^x$ denotes the parallel
transport from the starting vertex $s_{e}$ of $e$ to $x$.
The
phase space algebra is given by
\be \{X_{e}^a,g_{e}\}= g_{e}J^a;
\, \, \{X_{e}^a,X_{e}^b\}=-2{\e^{ab}}_{c} {X_{e}^c},
\ee  together with the relations
\be
g_{-e}=g_{e}^{-1}; \, \, X_{-e} = - g_{e}
X_{e}g_{e}^{-1},
\ee
 where $e$ denotes any edge and $-e$ denotes the reversed orientation.

Given
a face $f$ of $\G$ and a vertex $s_{f}$ of $f$ we define the
holonomy around $f$ starting from $s_{f}$,
$G(f)=\overleftarrow{\prod}_{\bar{e}\subset f}g_{\bar{e}}$.
We include only internal edges in the definition of $G(f)$,
and if $f$ contains an open end $v$ we chose $s_{f}$ to be the
terminal point of $e_{v}$.
We denote $G(f) = U(f)\mathrm{Id} + P^a(f) J_{a}$,
note that $U(f)^2 + P^a(f) P_{a}(f)=1$ since we
are in $\SU (2)$. We call $P(f)$ the momentum of $f$.
Given an internal edge $\bar{e}\subset f$ we can define
the `position' $Q_{\bar{e}}(f)$ and `orbital momenta' $L_{\bar{e}}(f)$
of $\bar{e}$ with respect to $f$ as follows.
First, lets introduce $\tilde{X}_{\bar{e}}(f)= G_{s_f}^{s_{\bar{e}}}(f)X_{\bar{e}}(G_{s_f}^{s_{\bar{e}}}(f))^{-1}$
where $G_{s_f}^{s_{\bar{e}}}(f)$ is the parallel transport from ${s_f}$ to $s_{\bar{e}}$ along $f$, then
\bea\displaystyle
Q_{\bar{e}}(f)&\equiv & U(f) \tilde{X}_{\bar{e}}(f) + \tilde{X}_{\bar{e}}(f)\wedge P(f) + \frac{\tilde{X}_{\bar{e}}(f)\cdot P(f)}{U(f)} P(f), \\
\displaystyle L_{\bar{e}}(f)&\equiv & -(P(f))^2 \tilde{X}_{\bar{e}}(f) +
U(f)\left(\tilde{X}_{\bar{e}}(f)\wedge P(f) + \frac{\tilde{X}_{\bar{e}}(f)\cdot P(f)}{U(f)} P(f)\right),
\eea
where $(X\wedge P)^a = {\e^a}_{bc} X^b P^c$ and $X\cdot X= X^aX_a$.
We can check that
\bea
\{Q^a_{\bar{e}}(f),P_b(f)\} = \delta^a_b,& &\,\, \{L^a_{\bar{e}}(f),P_b(f)\}= {\e^{a}}_{bc} P^c(f), \\
Q_{\bar{e}}(f)\wedge P(f) &=& L_{\bar{e}}(f).
\eea
Given any internal vertex $\bar{v}$ of $\Gamma$ we define
$J(\bar{v})= \sum_{e: s_{e}=\bar{v}} X_{e}$.

The constraints of the
theory are the flatness constraints and the Gauss law
\be
G(f) = 1;\ \ J(\bar{v})=0,
\ee
 for all internal
vertices $\bar{v}$ and all faces which do not contain open ends.
These constraints are first class.
For
faces $f_{v}$ which contains open ends $v$ we impose particle
constraints
\be
g_{e_{v}}^{-1}G(f_{v})g_{e_{v}} =h_{\t_{v}}; \, \, X_{e_{v}} = 2S_{v}J_{0}.
\ee
where $\theta_v = 4\pi m_v$ is half the deficit angle
created by the presence of a particle of mass $m_v$, $S_{v}$ is the spin of the particle
and $h_{\t_{v}}= \exp(\theta_v J_0)$.
Among these six constraints, four are second class and
two are first class, they generate time reparametrisation and $U(1)$
gauge symmetry. It is easy to check that
$P(f_{v})$ and $J({v}) \equiv -1/2 X_{-e_v}+ L_{\bar{e}}(f)$
commute with these constraints. The Dirac bracket involving these observables
is therefore equal to the original bracket and reads
\be
\{J^a(v),J^b(v)\} = {\e^{ab}}_{c} {J}^c(v), \, \,
\{J^a(v),P^b(f_{v})\} = {\e^{ab}}_{c} {P}^c(f_{v}).
\ee
In term of these observables the constraints are
\be\label{classspinconstraint} P^a(f_v) P_{a}(f_v) = \sin^2{\t_{v}};\ \ J_{a}(v)P^a(f_v) = {S_{v}}\sin{\t_{v}}.
\ee

The Gauss law generates gauge transformations
\be
\{J_\L, X_e\} = [X_e,\L(s_e)];\, \, \{J_\L, g_e\}= g_e \L(s_e) - \L(t_e) g_e,
\ee
where $\L(\bar{v})$ is a collection of Lie algebra elements
associated with internal vertices
and $ J_\L= \sum_{\bar{v}} J^a(\bar{v}) \L_a(\bar{v})$.
The momentum generates translational symmetry.
In order to understand
its action we choose a rooted maximal tree $T_R$ of $\G$ passing through
all internal  vertices, with a distinguished vertex $R$ (the root).
We denote by $G_{\bar{v}}^R(T)$ the parallel transport along the tree from any
internal vertex $\bar{v}$ to the root.
We choose a collection of Lie algebra element $\phi(f)$ for all faces of $\Gamma$.
These elements are all sitting at the root from the gauge transformation points of view.
We need to transport them to the faces using
 $G_{\bar{v}}^R(T)$.
Finally, we define
$P_\phi \equiv \sum_f P^a(f) [(G_{s_f}^R(T))^{-1} \phi(f) G_{s_f}^R(T)]_a$.
$P_\phi$ commutes with all group elements $g_e$, its commutation with
$X_e$ is given by
\be
\{P_{\phi},X_e \}= -\sum_{f\supset e} \e(f,e) \left(U(f) \tilde{\phi}_{s_e}(f) + {[} P_{s_e}(f), \tilde{\phi}_{s_e}(f) {]}\right).
\ee
where $\e(f,e)$ denotes the relative orientations of $f$ and $e$
and $$P_{s_e}(f)=(G_{s_e}^{s_f}(f))^{-1}P(f)G_{s_e}^{s_f}(f);\ \
\tilde{\phi}_{s_e}(f)=
\left(G_{s_f}^R(T)G_{s_e}^{s_f}(f)\right)^{-1}\phi(f)G_{s_f}^R(T)G_{s_e}^{s_f}(f),$$
$G(T),G(f)$ being the parallel transport along $T,f$.
This is the hamiltonian version of the translational
gauge symmetry defined in (\ref{eqn:discretetrans}).
We can now introduce a parametric time by taking $P_{\phi}$ to be our hamiltonian.
With respect to this time evolution all $J(\bar{v})$'s $X_{e_v}$ and $P(f)$'s are conserved.
However the $J(v)$'s or $X_e,Q_e(f)$ are not conserved and we could use one of them as a clock
in order to
define a relational time \cite{Waelbroeck:iz,Rovelli}.

\subsection{ Quantization}
The quantization of this system is implemented by taking the
spin network representation presented in section \ref{spinningpart}
and more precisely in section \ref{transpinpart}.
$\F_{(\G,j_{\bar{e}},I_v)}(g_{\bar{e}};g_{e_v}) \in \otimes_v V_{I_v}$
is a gauge invariant functional of all $g_{e}$'s valued in the tensor
product of representations associated with open ends.
The dependance on the open end group elements is
explicitly given by
\be \label{snorm}
\F_{(\G,j_{\bar{e}},I_v)}(g_{\bar{e}};g_{e_v}) = \otimes_{v}
D_{I_{v}}(g_{e_{v}}^{-1})\F_{(\G,j_{\bar{e}},I_v)}(g_{\bar{e}};1),
\ee
where $D_{I}(g)$ is the matrix element of $g$ in the representation $V_I$.

In the spin network representation the $g_e$'s are acting by  multiplication and the $X_e$'s are
acting by left invariant derivative
\be\label{opaction}
\hat{g}_e \F_{(\G)}(g_{e}) = {g}_e\F_{(\G)}(g_{e}) ;\,\, \hat{X}_e^a \F_{(\G)}(g_{e}) =
i\F_{(\G)}(g_{e}J^a).
\ee
The physical states are obtained  by `projecting' the spin network
states on the kernel of the constraints.
This `projector' is rigourously defined as a rigging
map \cite{Giulini:1998kf} mapping spin network states onto
distributional spin network states (Linear functionals on the space of
spin networks).
This rigging map is easily obtained: We have to multiply the spin
network states by a factor $\delta(G(f))$ for each face which does not
contain open ends.
We multiply by a factor $\delta(G(f_{v})g_{e_{v}}h_{\t_{v}}g_{e_{v}}^{-1})$
for each face with open ends.
Eventually we have to insert a spin projector $P_{I_{v},I_{v}}^{S_{v}}$
for each open end
where
$P_{I,I}^S = |S,I\ket\bra I,S|$, $|S,I\ket$ being the normalized vector
of spin  $S$ in
$V_I$, $\bra I,S|$ being its conjugate.
This is clearly a projector, moreover it satisfies
$P_{I_{v},I_{v}}^{S_{v}}D_{I_v}(J_{0}) =
-2iS_{v}P_{I_{v},I_{v}}^{S_{v}}$.
By an analysis similar to the one performed in eq. (\ref{pjs}), one can check that
\be
\hat{P}(f_v)^a \hat{J}(v)  P_{I_{v},I_{v}}^{S_{v}}\F_{(\G,j_{\bar{e}},I_v)}(g_{\bar{e}};g_{e_v})
=
S_v \sin\t_v P_{I_{v},I_{v}}^{S_{v}}\F_{(\G,j_{\bar{e}},I_v)}(g_{\bar{e}};g_{e_v}),
\ee
and leads to the imposition of the spin constraint
(\ref{classspinconstraint}) at the quantum level.
The physical scalar product between two spin network states is obtained by
integrating the product of the spin network states with the rigging map
over all group elements. It is explicitly written in section (\ref{transitionamp}).

One of the main goal is to be able to compute
the matrix elements of the unitary operator $\exp P_{\phi}$.
We are not there yet, but the Ponzano-Regge time observable we propose
is one step in this direction. More precisely, lets consider
$\S_{g,n}$  the genus $g$ surface with $n$ punctures. Using the gauge fixing
we can collapse  the triangulation of $\S_{g,n}$ to a triangulation with one vertex.

In this case the dual graph possess only one face and
$P_\phi =\tr \phi G_T$ where
$G_T= \prod_{i=1}^g [a_i,b_i]\prod_{p=1}^nu_ph_{\t_p}u_p^{-1}$.
Let us define the operator
\be
U_{|\phi|}= \int dg_R \exp({ iP_{g_R^{-1} \phi g_R}}).
\ee
In the appendix \ref{app:su2} we prove that
\be
\int \frac{d\phi}{4\pi} e^{i\tr(\phi g)} =
\int \frac{d|\phi|}{4\pi} |\phi|^2 U_{|\phi|}=
\sum_j dj^2 \frac{\chi_j(g)}{d_j}.
\ee
Therefore at the level of distributions
we have
\be
U_{|\phi|}= \delta(|\phi|-d_j)\frac{\chi_j(G_T)}{d_j}.
\ee
The inclusion of the time operator had as
an effect to replace the expectation value of $\delta(G_T)$ by
${\chi_j(G_T)}{d_j}$ ( see the definition of the time operator
and also section \ref{transitionamp}),
therefore it is computing the expectation value of the operator $U_{d_j}$.

%%%%%%%%%%%%%%%%%%%%%%%%%%%%%%%%%%%%%%%%%%%%%%%%%%%%%%%%%%%%%%%%%%%%%%%%%%%%%%%%%%%%
%%%%%%%%%%%%%%%%%%%%%%%%%%%%%%%%%%%%%%%%%%%%%%%%%%%%%%%%%%%%%%%%%%%%%%%%%%%%%%%%%%%%
%%%%%%%%%%%%%%%%%%%%%%%%%%%%%%%%%%%%%%%%%%%%%%%%%%%%%%%%%%%%%%%%%%%%%%%%%%%%%%%%%%%%
%%%%%%%%%%%%%%%%%%%%%%%%%%%%%%%%%%%%%%%%%%%%%%%%%%%%%%%%%%%%%%%%%%%%%%%%%%%%%%%%%%%%

\section{Massive spinless particles}\label{tranpartamp}
In this section we are going to describe the computation
of transition amplitudes for massive but non spinning particles.
\subsection{Definition of the amplitude}

We are considering a manifold $M$ with boundaries
\footnote{the boundaries are possibly empty}
$\partial M = \Sigma_i \coprod \Sigma_f$,
$\Delta$ a triangulation of $M$ and $\Delta^*$ the dual triangulation.
We denote by $\Gamma_i = \Delta^* \cap \Sigma_i$ and
$\Gamma_f = \Delta^* \cap \Sigma_f$ the closed trivalent oriented graph
obtained by the intersection of the 2-skeleton of $\Delta^*$ with
$\partial M$.
We label the edges $\bar{e}_i, \bar{e}_f$ of $\G_{i}, \G_f $ by spins $j_i,j_f$
and consider spin network functionals $\Phi_{\G_i,j_i}(g_{\bar{e}_i})$,
$\Phi_{\G_f,j_f}(g_{\bar{e}_f})$ defined in section (\ref{transamp}).

In our model, we consider that the particles propagate
along the edges of the triangulation.
In order to describe their propagation and interaction
we consider $\da$ an oriented graph with open ends
whose edges are edges of the triangulation $\D$, the open
ends of $\da$ all lie in the boundary and are identified with
vertices of the boundary triangulation $\D \cap \Sigma_{i,f}$.
The internal vertices of $\da$ describe the interaction of particles.
We label each edge $e$ of $\da $ by an angle $\theta_{e}$,
it is related to the mass of the particle by $\theta_{e}=4\pi G m_e$.
This correspond to a decoration
$D=(m_e,S_e;I_{s_e},I_{t_e},\imath_v)$ where $S_e=I_{s_e}=I_{t_e}=0$.
According to our gauge fixing procedure
we have to choose a maximal tree $T^*$ of $\D^*$ and a maximal
tree $T$ of $\Delta - \da$.

We define the transition amplitude to be
\be\label{spinlessamp}
\bra \P_{\G_f,j_f}|\P_{\G_i,j_i}\ket_{M,\da}=
\int \prod_{e^*\notin T^*} dg_{e^*}¥ \prod_{e \notin T\cup \da}\delta(G_e)
\prod_{e \in \da} \delta_{\theta_{e}}(G_e)\bar{\P}_{\G_f,j_f}(g_{\bar{e}_f})
{\P}_{\G_i,j_i}(g_{\bar{e}_i}).
\ee
This amplitude does not depend on the triangulation
$\D$ and the choices of trees.
It also depends only on the diffeomorphism class of $\da$
\cite{mathDL}.
The inclusion of particles amounts to insert a particle PR-observable
\be \label{spinlesspart}
\tO_{\da,\t_e}(j_e,g_{e^{*}})=\prod_{e\in\da} {\d_{j_e,0} \d_{\t_e}(G_e)}.\ee

Expanding the delta function
and integrating over $g_{e^*}$
we can write this amplitude in a Ponzano-Regge like
form\footnote{using recoupling theory and the expansion of
$\d_{\t_e}$  in terms of characters $\delta_{\t_e}(G)=\sum_j
\chi_j(h_{\t_{e}})\chi_j(G)$ (see appendix \ref{app:su2})}
\be
\bra \P_{\G_f,\j_f}|\P_{\G_i,j_i}\ket_{M,\da}=
\sum_{j_{e}} \left(\prod_{e\in T} \delta_{j_{e},0}\right)
\left(\prod_{e \notin T \cup \da}d_{j_{e}} \right)
\left(\prod_{e \in \da} \chi_{j_{e}}(h_{\t_{e}})\right)
\prod_{t}
\left\{
\begin{array}{ccc}
    j_{e_{t_{1}}} &  j_{e_{t_{2}}} &  j_{e_{t_{3}}} \\
    j_{e_{t_{4}}} &  j_{e_{t_{5}}} &  j_{e_{t_{6}}}
    \end{array}
    \right\},
    \ee
where the summation is over all spins belonging to
internal edges of $\D$, the spins of the boundary edges
$\bar{e}_{i}, \bar{e}_{f}$ given by $\j_{i},\j_{f}$ are not summed over and
$e_{t_{i}}$ denotes the six edges belonging to the tetrahedra $t$.

As we have seen in section \ref{Quamp}, we can
express the transition amplitude in the polarization were $e$ or $\o$ is fixed.
In order to change polarization we have to add a boundary term
$S_{b}= -\int_{\partial{M}} \tr(e\wedge \o)$.
The same is possible at the discrete level, we can express
the transition amplitude as a propagator $G_{M,\da}(g_{\bar{e}_i}, g_{\bar{e_f}})$
depending on the boundary connection:
\be
G_{M,\da}(g_{\bar{e}_i}, g_{\bar{e_f}})=\sum_{{\j}_i,\j_f}\left(\prod d_{j_i}d_{j_f}\right)
\bar{\P}_{\G_f,j_f}(g_{\bar{e}_f}){\P}_{\G_i,j_i}(g_{\bar{e}_i})
\bra \P_{\G_f,\j_f}|\P_{\G_i,j_i}\ket_{M,\da},
\ee
\be
\bra \P_{\G_f,\j_f}|\P_{\G_i,j_i}\ket_{M,\da}
= \int dg_{\bar{e}_i}dg_{\bar{e}_f} G_{M,\da}(g_{\bar{e}_i}, g_{\bar{e_f}})
\bar{\P}_{\G_f,j_f}(g_{\bar{e}_f}){\P}_{\G_i,j_i}(g_{\bar{e}_i}).
\ee
The term ${\P}_{\G,j}(g_{\bar{e}})$ is the discrete analog of
$\exp\int_{\partial M}\tr(e\wedge \o)$.

We are now going to give justifications for the amplitude \ref{spinlessamp}.
\subsection{Link with the discretized action }

At the  classical level the particle is inserted
by adding to the gravity action a particle term $\overline{S}_{P_{m}}= m \int
dt \tr(e_tJ_0)$.
The total action reads
$S_T= \int d^3x \tr\left(e(x)\w (F(\o) + 4Gm J_0 \delta_P(x))\right)$,
where $\delta_P(x)=\int dt \delta^{(3)}(x-x_P(t))$.
At the discrete level, the curvature term is replaced by the holonomy $G_e$ and the
frame field by the Lie algebra element $X_e$, we choose the discrete action to be
\be
S_{T}= \sum_{e\notin \da} \tr(X_eG_e) + \sum_{e\in\da}\tr(X_eG_eh_{\t_e}),
\ee
where $\t_e = 4\pi Gm_e$.
In the limit where the loops around $e$ become infinitesimal and the mass is small compare to the Planck mass
we recover the continuum action.
If we integrate over $X_e$ we get for each edge of $\da$ a factor
$\delta(G_eh_{\t_e})$.
This is not yet the term we want, we have to remember that when we introduce a particle
via the action $\overline{S}_{P_{m}}$ we are breaking the gauge symmetry at the location of the particle.
In order to define $G_e$ we have to choose a dual vertex $st(e)$ along the face dual to $e$,
the discrete Lorentz gauge group is acting at this vertex $G_e \rightarrow u_e^{-1}G_eu_e$.
If we integrate over the gauge group action
we get
\be
\int du_e \delta(G_eu_eh_{\t_e}u_e^{-1}) = \delta_{\t_{e}}(G_e).
\ee
So the insertion of a particle forces $G_e$ to be in the same conjugacy class as
$h_{\t_e}$.
Note that now the momentum of the particle is
no longer labelled by a Lie algebra element as in (\ref{p}), but by
a group element $u_eh_{\t_e}u_e^{-1}$.
This property is characteristic of a theory exhibiting
the symmetry of Doubly Special Relativity \cite{FLeeJerzy}.
The fact that a spinless particle coupled to gravity can be inserted
by imposing the holonomy around the particle to be in a given conjugacy class
can be easily seen from our hamiltonian analysis.
It
was first recognized, at the classical level, in \cite{Waelbroeck:dt}
and analyzed in great details in \cite{Matschull:1997du}.
Our amplitudes extend these works to the quantum case.

\subsection{Physical interpretation of the particle observables}\label{phyint}

The partition function contains for each edge $e$ of the
triangulation a term $\d(G_e)$ which ensures that the curvature
around $e$ is zero. The insertion of a PR-observable $\tO_{\da,\t_e}$
removes the zero curvature condition and replaces it by the fact
that the curvature around $e$ has to be in the conjugacy class of
$h_{\t_e}$. This can be interpreted as the presence of a particle
of mass $m_e=\t_e/4\pi G $ moving along $e$, creating a line of conical
singularity. For such an interpretation to
be valid, we have to check that it satisfies a kind of 'mass
conservation property'.
\begin{figure}[ht]
\includegraphics[width=3cm]{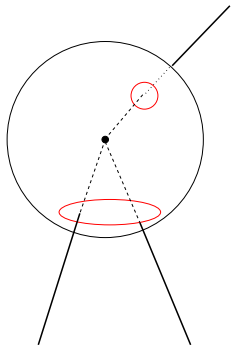}
\caption{particle vertex}\label{bianch3particles}
\end{figure}

Consider a bivalent vertex of the graph. The two edges of this
graph carry group elements $g_1$ and $g_2$ which are imposed to
lie in the conjugacy class labelled by, say, $\t_1$ and $\t_2$. In
the triangulation this vertex is surrounded by other edges, with
group elements $g_e$ which are imposed to be $1$ due to the
flatness condition. Now the overall still satisfies the Bianchi
identity, which in this case reduces on-shell to $g_1g_2^{-1}=1$ (provided
we choose an ingoing orientation for the first edge and an
outgoing for the second one). This proves that the expectation
value of the particle PR-observable is zero unless $\t_1=\t_2$.
If fact, if there is a bivalent vertex, this analysis shows that
the amplitude (\ref{spinlessamp}) will be proportional
to $\delta_{\t_1}(h_{\t_2})$ in this case the
physical amplitude is the proportionality coefficient.
If we naively put equal masses at bivalent vertices in
(\ref{spinlessamp}) we get an infinite result, this is due to the fact that there is
an extra gauge symmetry associated with the propagation of a particle which is time
reparametrisation.

Consider now a trivalent vertex. This time the Bianchi identity
around the vertex  reduces to
\begin{equation}
g_1=g_2g_3 \mbox{ or }
u_1h_{\t_1}u_1^{-1}=u_2h_{\t_2}u_2^{-1}u_3h_{\t_3}u_3^{-1}.
\end{equation}
This can only holds if $m_1$,$m_2$ and $m_3$ satisfy the
inequality
\begin{equation}
\cos(\frac{\t_2+\t_3}{2})\leq \cos\frac{\t_1}{2}\leq
\cos(\frac{\t_2-\t_3}{2}).
\end{equation}
This inequality is the analog of the kinematical inequalities of
relativistic particles. It is shown in the companion paper \cite{mathDL} that
this condition actually corresponds to the existence of the
intertwiner between the associated three representations of
$DSU(2)$.

\section{Massive Spinning Particles}\label{transpinpart}
In this section we are going to describe the inclusion of
quantum massive spinning particles coupled to quantum gravity.

\begin{figure}[ht]\psfrag{vext}{$v_{i}$}\psfrag{vint}{$\bar{v}_{i}$}
\psfrag{jv}{$e_v$}
\includegraphics[width=4cm]{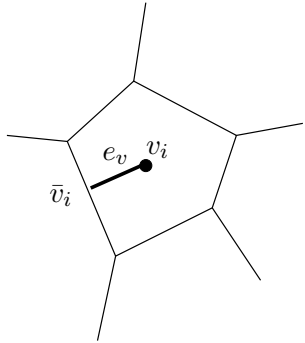}
\caption{particle vertex}\label{openend}
\end{figure}

We are considering a manifold $M$ with boundaries\footnote{the boundaries are possibly empty}
$\partial M = \Sigma_i \coprod \Sigma_f$,
$\Delta$ a triangulation of $M$ and $\Delta^*$ the dual triangulation.
We denote by $\Delta^*_i = \Delta^* \cap \Sigma_i$ and
$\Delta^*_f = \Delta^* \cap \Sigma_f$ the closed trivalent oriented graph
obtained by the intersection of the 2-skeleton of $\Delta^*$ with
$\partial M$.
We denote by $v_i$ (resp. $v_f$) the vertices of $\D \cap \Sigma_i$ (resp. $\D \cap \Sigma_f$).
Each $v_i$ is at the center of a face of $\Delta^*_i$, for each $v_i$ we choose
a vertex $\bar{v}_{i}$ lying in the middle of one of the dual edges  surrounding
$v_i$ and draw an edge called ${e}_{{v}_i}$ from $v_{i}$ to $\bar{v}_{i}$,
see figure (\ref{openend}).
We denote by $ \Gamma_i = \Delta^*_i \coprod \{{e}_{v_i}\}_i $ the union of
$\Delta^*_i$ with these edges, this is an oriented graph with open ends.
We do the same construction for $ \Sigma_f$ and denote the resulting graph $\Gamma_f$.
We color the internal edges $\bar{e}_{i}$ of $\Gamma_i$ by $SU(2)$ spins $j_{\bar{e}_i}$ and the
open edges  ${e}_{{v}_i}$ of $\Gamma_i$ by $SU(2)$ spins $ I_{{v}_i}$.
The resulting spin network with open ends is denoted
$(\Gamma_i, j_{\bar{e}_{_i}},I_{{v}_i})$ and similarly for $\Gamma^*_f$.

Given such data we construct a spin network functional denoted
by $\F_{(\G^*_i,j_{\bar{e}_i},I_{{v}_i})}$ which belongs to
\be
{\cal{H}}_{\Gamma,I_v} = L^{2}\left((G^{|{e}|}\rightarrow \otimes_v V_{I_v}) /G^{|\bar{v}|}\right),
\ee
where $|{e}|$ denotes the number of  edges of $\Gamma$,$|\bar{v}|$ the number of internal vertices
of $\G$,
  $v$ the  open ends of $\Gamma^*$,
and the action of $G^{|\bar{v}|}$ on $G^{|{e}|}$ is given by the structure of the graph.
The functional is explicitly defined as
\be
\F_{(\G,j_{\bar{e}},I_v)}(g_{\bar{e}};g_{e_v}) =
\bra \otimes_{\bar{v}} C_{\bar{v}} |\otimes_{\bar{e}} D^{j_{\bar{e}}}(g_{\bar{e}})
\otimes_{v} D^{I_{v}}(g_{\bar{e}_v})\ket \in \otimes_v \left(V_{I_v}\right)^G,
\ee
where $C_{\bar{v}} $ are normalized intertwiners for  internal vertices.
This is a  construction similar to the one presented in  section \ref{Quamp}, but adapted to
a triangulation.

These spin network states arise naturally in the hamiltonian
quantization of 3D gravity with spinning particles as kinematical states
satisfying the Gauss law.
The operators $g_{{e}}$ act by multiplication and the operators $X_{e}$ as left invariant derivatives,
see (\ref{opaction}).

\subsection{Particle graph functional}

The purpose of this section is to describe the construction
of a particle graph functional which contains the data
characterizing the propagation and interaction of
particles in the Ponzano-Regge model. The data we need are the following:
we consider $\da$ an oriented graph with open ends
whose edges are edges of the triangulation $\D$, the open
ends of $\da$ all lie in the boundary and are identified with the boundary
vertices $v_i, v_f$.

Each edge $e$ of $\da$ is labelled by a mass and a spin $(m_e, S_e)$;
also each starting point of an edge $e$ is denoted by $s_e$ and labelled by a $\SU (2)$ representation
$V_{I_{s_e}}$, each terminal point of an edge $e$ is denoted by $t_e$ and labelled by a
$\SU (2)$ representation $V_{I_{t_e}}$, both $I_{s_e}-S_e, I_{t_e}-S_e$ are positive integers
and each internal vertex $\tilde{v}$ of $\da$ is labelled by an intertwiner
$\imath_{\tilde{v}}$.
The collection of data $(m_e,S_e;I_{s_e},I_{t_e},\imath_{\tilde{v}})$ is called a decoration
and denoted for short $D$, the decorated particle graph is
denoted $\da_D$.

\subsubsection{Spin projector}
In order to construct the particle graph functional we need to
introduce an important object: {\it the spin projector} $\Pi_{I,I'}^s :G\rightarrow V_I\otimes V_{I'}$
which plays a key role in the construction of the propagator
of spinning particles coupled to gravity
\be
\Pi_{I,I'}^s(u) \equiv D^I(u^{-1}) P_{I,I'}^s D_{I'}(u),
\ee
where $D^I(u)$ is the matrix element of $u$ in the representation $V_I$ and
$P_{I,I'}^s = |s,I\ket\bra I',s|$, $|s,I\ket$ being the normalized vector of spin  $s$ in
$V_I$, $\bra I,s|$ being its conjugate.
It is clear that $I-s, I'-s' \in \N$ in order for this definition to make sense.
The spin projector is characterized by the following properties
\bea
\Pi_{I,I'}^s(u)\Pi_{I',I''}^{s'}(u) &=& \delta_{s,s'} \Pi_{I,I''}^s(u),\\
\Pi_{I,I'}^s(hu)&= &\Pi_{I,I'}^s(u),\mathrm{for\ all}\ h \in H,\\
\Pi_{I,I'}^s(gu) & = & D^I(g^{-1})\Pi_{I,I'}^s(u)D^{I'}(g) \mathrm{for\ all}\ g \in G,\\
\Pi_{I,I'}^s(1)&=& P_{I,I'}^s.
\eea
The first property justifies the name projector for $\Pi$, the second
property shows that $\Pi$ is a functional on $H\backslash G$.
The space $H\backslash G$ is isomorphic to the momenta space of a particle:
In the Lorentzian case, the momenta $p \in \R^3$ of a 3d particle lie in the upper sheet
${\cal{H}}\simeq {\SO} (2,1)/{\SO} (1,1)$ of the hyperboloid, due to the mass shell condition
$p^2=m^2$.
Given $p$ we can choose a section $u:{\cal{H}} \rightarrow G$. If a functional $f(u(p))$ is invariant
under the right action of $H$ then $f$ doesn't depend on the choice of the section and is
therefore a functional of the momenta.
In our case the space of Euclidean momenta is the sphere, and $u$ is interpreted as
a particle momenta.
The third property shows that $\Pi$  intertwines  the action of
boosts on the momenta and on the external Lorentz index.
The last property is a normalization condition showing that we are describing a spin $s$
particle.
Note that $\Pi$ does not depend on the mass of the particle.

One of the main justification for introducing $\Pi$ is that it appears
naturally in the computation of the Feynman propagator of spin $s$ particle
propagating in flat space.
Let $\phi^s(x)$ be a field of spin $s$, which is a function valued in $V_s$ and let us denote
by $\partial^{(n)}\phi^s(x)$ the traceless $n$-th derivative of $\phi^s$, it is a function valued
in $V_n\otimes V_s$ since $\partial^{(n)}$ is specified once we choose a traceless symmetric product
of vectors in $V_1$. We can project $V_n\otimes V_s$ onto  $V_{n+s}$ and denote by
$\partial^{(n}\phi^{s)}(x)$ the corresponding field.
The flat space Feynman propagator for this field is
\be
\bra\partial^{(I-s}\phi^{s)}\partial^{(I'-s}\phi^{s)}\ket(p)
= \frac{\Pi_{I,I'}^s(u(p))}{p^2-m^2+i\epsilon}.
\ee
This is proven in \cite{FLL} in the three dimensional case,
following the analogous proof of \cite{Weinberg} in the case of dimension $4$.

\subsubsection{Framed particle graphs}
\begin{figure}[ht]\psfrag{te}{$t_e $}\psfrag{fe}{$f_e^*$}
\psfrag{ste}{$st(e)$} \psfrag{se}{$s_e$}\psfrag{gf}{}
\includegraphics[width=2cm]{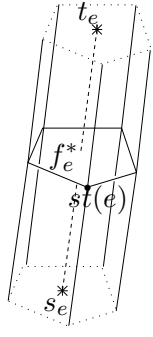}
\caption{tube surrounding a particle worldline}\label{tube}
\end{figure}

\begin{figure}[ht]\psfrag{te}{$t_e $}
\psfrag{ste}{$st(e)$} \psfrag{se}{$\!s_e$}\psfrag{gf}{}
\includegraphics[width=8cm]{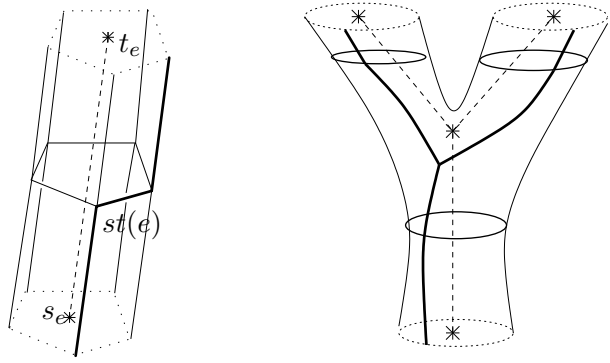}
\caption{The framed graph}\label{framed}
\end{figure}

In order to construct the particle graph functional we need to give a {\it framing}
to $\da$.
We associate to $\da \in \D$ the set of
edges of $\D$ which touche $\da$ but are not in $\da$
\be
T_\da = \{ e\in \D_1\ |\ e\cap \da \neq \emptyset \ \mathrm{and} \  e\notin \da  \},
\ee
where $\D_1$ refers to the one-skeleton of $\D$.
Each edge of $e\in\D$ can be viewed as a dual face $f^*_e$ of $ \D^*$.
We define the {\it tube} of $\da$, denoted $ T^*_\da$ to be the set of dual faces corresponding to
edges in $T_\da$. So $e \in T_\da $ iff $ f^*_e \in T^*_\da$.
The neighborhood of $\da$ from the point of view of $\D^*$ is given by
$T^*_\da$ together with the set of dual faces $f_e^*,\ e \in \da$.
$T^*_\da$ is made up of tubes or cylinders surrounding the edges of $\da$
and the topology of $T^*_\da$ is a 2d surface with holes which
can be viewed as the boundary of a thickening of $\da$, the holes
being at the location of the particle (see figure \ref{tube}).
We denote by $\da^f$ and call it  a  framed particle graph, a graph in $\D^*$
with the following properties:
$\da^f$ is a graph  entirely lying in $T^*_\da$, it is
topologically equivalent to $\da$ and provides $\da$ with a framing $f$.
The end points of $\da^f$ are identified with the boundary
dual vertices $\bar{v}_i,\bar{v}_f$ introduced in the previous section.
The framing being given by a choice of vectors on $\da$ pointing towards $\da^f$.
We induce the decoration of $\da$ on $\da^f$ and denote the corresponding
decorated graph by $\da^f_D$, see figure (\ref{framed}).
For each vertex $v$ of $ \da$ we choose a corresponding vertex
$v^*$ of $ \da^f$, however not all vertices of $\da^f$ are of these type
there is for each edge of $\da$ a vertex $st(e)$ which comes from the intersection
of $\da^f$ with $f^*_e$, (see figure \ref{framed}).
\begin{figure}[ht]\psfrag{vext}{$v $}\psfrag{Gf}{$\da^f$}
\psfrag{G}{$\da$}
\includegraphics[width=3cm]{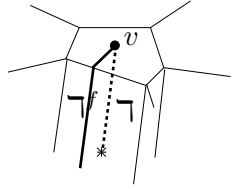}
\caption{connection between the framed graph and the boundary spin network}\label{boundvertex}
\end{figure}
Given an edge $e$ of $\da_D$ we consider
\be
\Pi_{I_{t_e},I_{s_e}}^{S_e}(g_e^*,u_e) =
D_{I_{t_e}}((g^*_{t_e})^{-1})\Pi_{I,I'}^s(u_{e}¥) D_{I_{s_e}}(g^*_{s_e})
\in \mathrm{Hom}( V_{I_{s_e}}, V_{I_{t_e}}),
\ee
where $g^*_{s_e}$ is the group element corresponding to the holonomy
along the dual edges of $\da^f$  going
from $s_e^*$ to $st(e)$ and $g^*_{t_e}$ corresponds to the holonomy
from  $t_e^*$ to $st(e)$ .

We can finally define the particle graph functional
to be
\be\label{partgfunc}
\Pi_{\da^f_D}(g_{e^*},u_e) =
\bra \otimes_{{v} \in \da} \imath_{{v}} |\otimes_{\bar{e}\in \da}
\Pi_{I_{s_e},I_{t_e}}^{S_e}(g_e^*,u_e)\ket
 \in \mathrm{Hom}\left(\otimes_{v_i} V_{I_{v_i}}; \otimes_{v_f} V_{I_{v_f}}\right).
\ee
The PR-observable that corresponds to the inclusion of interacting spinning particles
is given by
\be\label{spinPRob}
\Pi_{\da^f_D}(j_e,g_{e^*})=\prod_{e\in \da} \delta_{j_e,0}
\int \left(\prod_{e\in \da}du_e \delta(G_e u_e h_{\t_e}u_e^{-1})\right) \Pi_{\da^f_D}(g_{e^*},u_e).
\ee
This operator is clearly a generalization of the operators we have already considered.
For instance if we take the decoration such that
all $S_e$ and all $I_v$ are zero we recover the operator (\ref{spinlesspart}) describing
the propagation of a spinless particle without external angular momenta.
If we take a decoration such that ${\t_e}=0$, we chose $I_{s_e}= I_{t_e}$ and we sum
over $S_e$ we recover the Wilson line operator (\ref{wilsonline}).

In the case of spinning particles, the spin network states are functional
of the holonomies ${g}_{\bar{e}}$ along internal edges and of the holonomies
$(g_{\bar{e}_v})$ along the external edges.
$\F_{(\G_i,j_{\bar{e}_i},I_{v_i})}(g_{\bar{e}};g_{\bar{e}_v}) $
(resp. $\F_{(\G_f,j_{\bar{e}_f},I_{v_f})}(g_{\bar{e}};g_{\bar{e}_v})$)
are valued in $\otimes_{v_i} V_{I_{v_i}}$ (resp $\otimes_{v_f} V_{I_{v_f}}$).
We can therefore contract these functionals (we denote them $\F_i,\F_f$ for short)
 with the particle graph functional (\ref{partgfunc}) to get a scalar function
 \be
\bra \F_{f}(g_{\bar{e}};g_{\bar{e}_v}))|\Pi_{\da^f_D}(j_e,g_{e^*}) |
\F_{i}(g_{\bar{e}};g_{\bar{e}_v})\ket.
\ee
The transition amplitude is given by the following integral
\be\label{partgfuncamp}
\bra \F_{f}| \F_{i}\ket_{\da^f_D} =
\sum_{j_e|e\notin T} \int\prod_{e^*\notin T^*|e^*\neq \bar{e}_v} dg_{e^*}
\bra \F_{f}(g_{\bar{e}};1)|\Pi_{\da^f_D}(j_e,g_{e^*}) |
\F_{i}(g_{\bar{e}};1)\ket.
\ee
Note that we integrate only over the internal  boundary variables $g_{\bar{e}}$ and we fix the
value of $g_{\bar{e}_v}$ to be unity.
This amplitude depends on the choice of the framing. If we make a change of framing
along the edge $e$ by adding one more twist one sees that the amplitude is modified by a multiplicative
factor $ e^{i\theta_v 2S_v}$.

\subsection{Spin conservation }
It is now easy to check that the expectation value of the PR-observable
(\ref{spinPRob}) satisfies a mass and a spin conservation property.
The argument is similar to the one presented in section \ref{phyint}.
Lets consider a bivalent vertex $v=t_{e_{1}}=s_{e_{2}}$, this vertex
is surrounded by edges whose group elements $G_{e}$ are imposed to be
unity by the flatness constraint. Therefore, the Bianchy identity reduces to
\be
(g^*_{t_{e_{1}}})^{-1}G_{e_{1}}g^*_{t_{e_{1}}} =
(g^*_{s_{e_{2}}})^{-1}G_{e_{2}}g^*_{s_{e_{2}}},
\ee
where
$g^*_{t_{e_{1}}}, g^*_{s_{e_{2}}}$ corresponds to the holonomy
from  $v$ to $st(e_{1}), st(e_{2})$.
$G_{e_{1}}$ is imposed to be equal to $ u_{e_{1}}^{-1}h_{-\t_{e_{1}}}u_{e_{1}}$
and similarly for $G_{e_{2}}$. This implies that $h_{-\t_{e_{1}}}$ is
conjugated to $h_{-\t_{e_{2}}}$, so $ \t_{e_{1}}=\pm \t_{e_{2}}$.
This is the mass conservation.
If the sign is $+$, this means that $ u_{e_{1}}g^*_{t_{e_{1}}}=
h u_{e_{2}}g^*_{s_{e_{2}}}$ where $h$ is an arbitrary Cartan group element.
The PR observable (\ref{spinPRob}) contains the product
\be
D_{I_{t_{e_{2}}}}((g^*_{t_{e_{2}}})^{-1})\Pi_{I,I'}^{S_{1}}¥(u_{e_{2}}¥)
D_{I_{s_{e_{2}}}}(g^*_{s_{e_{2}}})
D_{I_{t_{e_{1}}}}((g^*_{t_{e_{1}}})^{-1})\Pi_{I',I''}^{S_{2}}(u_{e_{1}}¥)
D_{I_{s_{e_{1}}}}(g^*_{s_{e_{1}}}),
\ee
which is equal to
\be
\delta_{S_{1},S_{2}}
D_{I_{t_{e_{2}}}}((hu_{e_{2}}g^*_{t_{e_{2}}})^{-1})
P_{I,I''}^{S_{1}}
D_{I_{s_{e_{1}}}}(u_{e_{1}}g^*_{s_{e_{1}}}),
\ee
and the spin is conserved.

If $\t_{1}=-\t_{2}$ then $ u_{e_{1}}g^*_{t_{e_{1}}}=
\e h u_{e_{2}}g^*_{s_{e_{2}}}$ where $h$ is an arbitrary Cartan group element
and $\e$ is the Weyl group element
$$\e =\left(
\begin{array}{cc}
  0 & -1 \\
 1 & 0
\end{array}
\right). $$
  When acting on $|I,S\ket$, $D_{I}(\e)$ gives $|I,-S\ket$.
  We see that in this case the
  expectation value of (\ref{spinPRob}) contains a factor
  $\delta_{S_{1},-S_{2}} $ and the absolute value of the spin is conserved.
  This is the quantum analog of the remark at the end of section
  \ref{clapar}, showing that we have to include at once both
  spins $S$ and $-S$.
  If we have two incoming particles and one outgoing at an
  interaction vertex the same reasoning implies that
  $|S_{1}|+ |S_{2}|= |S_{3}|$

\section{Computation of transition amplitudes}\label{transitionamp}

In this section we present explicit computations of transition amplitudes
(\ref{spinlessamp},\ref{partgfuncamp}) in the simple cases
where we have non interacting spinless and spinning particles on a genus $g$ surface.
We show that our general definition is free of any infinities in this case and reproduces
the physical scalar product one expects from the canonical quantization.
We also compute explicitly the effect of the exchange of two particles
and show that the result is characterized by the braiding matrix of the quantum group
 $D_\kappa(\SU(2))$.
 We also discuss the finiteness of our amplitudes for closed manifold.

\subsection{The physical scalar product and time propagation}
\begin{figure}[ht]
\includegraphics[width=12cm]{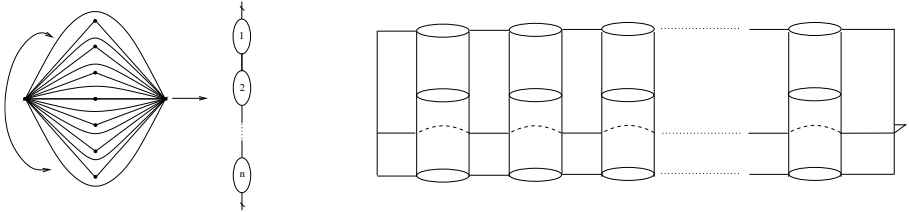}
\caption{Triangulation and dual triangulation of $\S_{0,n}$
and dual triangulation of $\S_{0,n} \times I$}\label{Sntriang}
\end{figure}

We are now going to specialize to the case of non-interacting particles.
We consider a simple triangulation of $\Sigma_{0,n}$, the sphere minus n points,
in terms of $2n$ triangles and $n+2$ vertices.
The triangulation and its dual are drawn in figure \ref{Sntriang}.
The particles are sitting inside the small tubes.

\begin{figure}[ht]\psfrag{g1}{$\bar{g}_i$} \psfrag{g2}{$\bar{g}_f$} \psfrag{kl}{$k^L$} \psfrag{kr}{$k^R$}
\includegraphics[width=7cm]{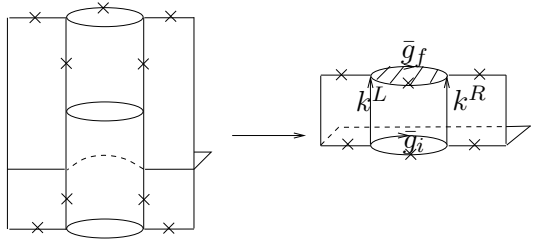}
\caption{gauge fixing around a tube: The crossed edges are gauge fixed.}\label{tubegaugefix}
\end{figure}

We consider a triangulation $\D$ of $\S_{0,n}\times I-\da_{n}$,
where $\da_{n}$ is the graph consisting of $n$ vertical unbraided
segments $\{x_i\}\times I$.
The triangulation we take contains $6n$ tetrahedra with all its vertices
 in the boundary, the dual triangulation is drawn in
figure \ref{Sntriang}.
We now chose a gauge fixing condition. Since all triangulation vertices are
in the boundary we need to gauge fix only the Lorentz symmetry.
In figure \ref{tubegaugefix} we  crossed the gauge fixed edges and show in the RHS the result of the gauge fixing.
We can simplify the picture by using the flatness condition
around dual faces where all edges but one are gauge fixed to eliminate
the last edge of this face. After this procedure we get a simpler
picture drawn in the RHS of fig. \ref{tubegaugefix}
which contains $4n+2$ non gauge fixed edges and $4n+2$ faces.
We label as in the figure the group elements around a tube $p$
by $k_p^L, k_p^R, \bar{g}_{p,i}, \bar{g}_{p,f}$ where the k's
label internal edges and we integrate over them.
we also denote by $\bar{g}_{0,i}, \bar{g}_{0,f}$ the edges of the boundary going from
the tube $n$ to the tube $1$.
The two vertical faces lying on the tube give us two delta functions:
$\delta(k_p^L(k_p^R)^{-1}) \delta(k_p^L\bar{g}_{p,f}(k_p^R)^{-1}(\bar{g}_{p,i})^{-1})$,
the horizontal face closing the tube gives us $ \delta_{\theta_p}(\bar{g}_{p,f})$,
the faces going from the tube $p$ to $p+1$, $p=1,\cdots, n-1$
give $\delta( k_p^R(k_{p+1}^L)^{-1})$,
the face going from the tube
$n$ to $1$ gives $\delta( k_n^R\bar{g}_{0,f}(k_{1}^L)^{-1}\bar{g}_{0,i}^{-1})$,
eventually, there are two faces encircling all the tubes one gives
$\delta(\bar{g}_{0,i}\bar{g}_{1,i}\cdots \bar{g}_{n,i})$,
the other one gives $\delta(\bar{g}_{0,f})$.
After integration over all $k$'s variables but one, we are left with the propagator
\be
G_n(\bar{g}_{p,f},\bar{g}_{p,i})=
\int dk \left(\prod_{p=1}^n \delta(k\bar{g}_{p,f}k^{-1}(\bar{g}_{p,i})^{-1})
\delta_{\theta_p}(\bar{g}_{p,f})\right)
\delta(\bar{g}_{1,i}\cdots \bar{g}_{n,i})\delta(\bar{g}_{0,f})\delta(\bar{g}_{0,i}).
\ee
After gauge fixing and taking into account the flatness condition for $\bar{g}_0$,
the boundary spin network states depend only on
$(\bar{g}_p)_{p=1,\cdots, n}$.
The physical scalar product between two spin network states is
then simply
\be\label{nprod}
\bra\Phi_f|\Phi_i\ket =
\int \prod_{p=1}^n du_p \bar{\Phi}_f(u_p h_{\t_p}u_p^{-1}){\Phi}_i(u_p h_{\t_p}u_p^{-1})
\delta(u_1 h_{\t_1}u_1^{-1}\cdots u_n h_{\t_n}u_n^{-1}).
\ee
This shows that the physical Hilbert space of $n$ particles on a sphere
 is isomorphic to $L^2(\SU(2)/U(1))^{n-1}$ and a physical state
 $\Phi_P(u_1,\cdot, u_{n-1})$ can be viewed as a distributional
 functional $\tilde{\Phi}$
 on the kinematical Hilbert space using a version of the so called `rigging map' \cite{Giulini:1998kf}
 \be
 \tilde{\Phi}_P(\bar{g}_p)= \delta(\bar{g}_1\cdots \bar{g}_n)\int
 \left(\prod_{p=1}^n du_p  \delta(\bar{g}_p u_ph_{\t_p}(u_p)^{-1})\right)
\Phi_P(u_1,\cdots, u_{n-1}).
\ee
We recover in this way the scalar product that we obtained in the Hamiltonian quantization
in section \ref{Hamfor}.
Also the physical Hilbert for $n$ particles coupled to gravity
is isomorphic to the Hilbert space of $n$ particles on the sphere
without gravity and can be written as the tensor product of
$n-1$ Poincar\'e representations $ \bigotimes_p {\cal H}_{\t_p,0}$.
However, even if the Hilbert spaces are isomorphic the representation of operators acting on
it is very different, this is exemplified in the next section where we compute the action of the braiding
operators.

If we insert the `time' operator (\ref{Time}) in the computation of the amplitude
its effect is to replace the $\delta$ function by a character $\chi_j$ and we get
\be\label{nprodtime}
\bra\Phi_f|\Phi_i\ket_j =
\int \left(\prod_{p=1}^n du_p \bar{\Phi}_f(u_p h_{\t_p}u_p^{-1}){\Phi}_i(u_p h_{\t_p}u_p^{-1})\right)
\chi_j(u_1 h_{\t_1}u_1^{-1}\cdots u_n h_{\t_n}u_n^{-1}).
\ee

\subsection{Spinning amplitudes}
We have seen that the amplitude including
spinning particles is given by
\be
\bra \F_{f}| \F_{i}\ket_{\da^f_D} =
\sum_{j_e|e\notin T} \int\prod_{e^*\notin T^*|e^*\neq \bar{e}_v} dg_{e^*}
\bra \F_{f}(g_{\bar{e}};1)|\Pi_{\da^f_D}(j_e,g_{e^*}) |
\F_{i}(g_{\bar{e}};1)\ket.
\ee
We want to specialize to the case where we have $n$ non-interacting spinning particles on the sphere
using the triangulation described in the previous section.
In this case the spin network states are functional
of the holonomies $(\bar{g}_p)_{p=1,\cdots n}$ around the particle and the holonomies
$(u_p)_{p=1,\cdots n}$ along the external edges.
$\F_{i}(\bar{g}_p;u_p) $ (resp. $\F_{f}(\bar{g}_p;u_p)$)
are valued in $\otimes_{v_i} V_{I_{v_i}}$ (resp $\otimes_{v_f} V_{I_{v_f}}$).
For non interacting particles the particle graph functional
factorises as a product of projectors.
It is  therefore easy to contract the spin network functionals
 with the particle graph functional (\ref{partgfunc}).
 The computation of the amplitude is similar to the one made in the previous section,
 if we restrict to the case where $\F_{i}=\F_{f}$ it reads
\be
\bra \F_{f}| \F_{i}\ket_{\da^f_D} = \int\prod_{ v } du_v
|\left\bra \F_{f}(u_vh_{\t_v} u_v^{-1};u_v)|\otimes_v |I_v,S_v\ket\right\ket|^2.
\ee

This shows that the physical Hilbert space of $n$ particle on a sphere
 is isomorphic to $\otimes_v{\cal H}_{S_v}$
where
\be
{\cal H}_S=\{ F\in L^2(G)| \, F(h_\theta g) =e^{i2\theta S}F(g)| \forall h_\theta \in H\}.
\ee
A physical state
 $\Phi_P(u_1,\cdot, u_{n-1})\in \otimes_v{\cal H}_{S_v}$ can be viewed as a distributional
 functional $\tilde{\Phi}\in \otimes_v V_{I_v}$
 on the kinematical Hilbert space
 \be
 \tilde{\Phi}_P(\bar{g}_v; u_v)= \delta(\bar{g}_1\cdots \bar{g}_n)
 \left(\prod_{v=1}^n du_v  \delta(\bar{g}_v u_v h_{\t_v}(u_v)^{-1})\right)
 \left( \otimes_v | I_v,S_v \ket \right)
\Phi_P(u_1,\cdots, u_{n-1}).
\ee

\subsection{ Braiding of particles and  general surface}

Using these techniques it is possible to compute the expectation value of a more general
particle graph. If we take $M$ to be the three sphere and consider a closed decorated particle graph
$\da_{ D}$. It is shown in \cite{mathDL} that the PR evaluation $Z(M,\da_{D})$ is equal to a
quantum group evaluation using $D_\kappa(\SU(2))$ a kappa deformation of the Poincar\'e group.
We are now going to show this explicitly by looking at the  exchange two particles.
The effect of gravity is to give a non trivial statistic to the particles
which is governed by a deformation of the Poincar\'e group.
We will explicitly show this now.

\begin{figure}[ht]\psfrag{k}{$k$}\psfrag{g1}{$\!\bar{g}_{1i}$}\psfrag{g'1}{$\bar{g}_{1f}$}
\psfrag{g2}{$\bar{g}_{2i}$}
\psfrag{g'2}{$\bar{g}_{2f}$}
\includegraphics[width=10cm]{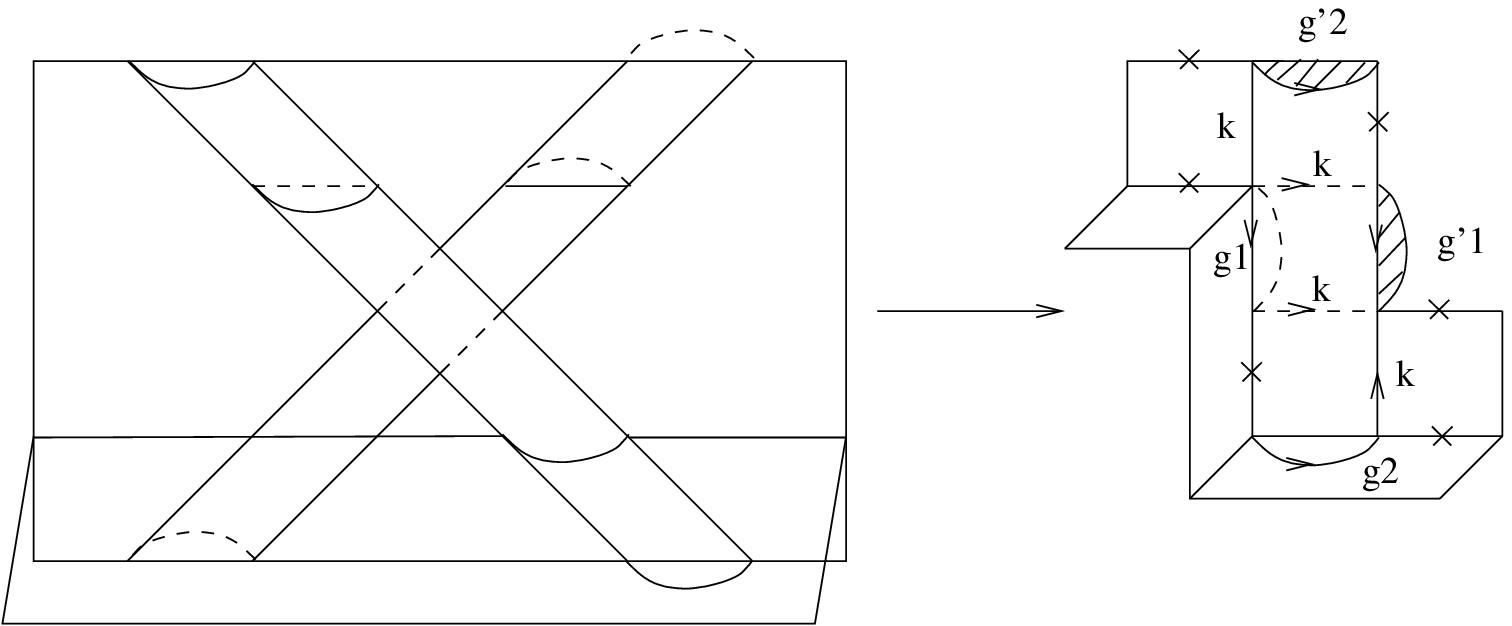}
\caption{dual triangulation of $\S_{0,n}\times I-\da_{n}$}\label{2exchange}
\end{figure}

We consider a triangulation $\D$ of $\S_{0,n}\times I-\da_{n}$,
where $\da_{n}$ is the graph consisting of $n$ vertical
segments. $n-2$ of them are unbraided and the segment 1 is crossing the segment two
so that the particles are exchanged.
The triangulation we consider is the same as the previous one except around the particles 1 and 2.
The corresponding dual complex is drawn in the LHS of figure (\ref{2exchange}).
Following the same path as previously we can gauge fixed the amplitude and collapse
the complex to the LHS of figure (\ref{2exchange}) where the gauge fixed edges are crossed.
Only four faces of the collapsed dual triangulation contribute non trivially,
its evaluation gives
\be
\int dk \delta(k\bar{g}_{1,f}\bar{g}_{2,f}k^{-1}\bar{g}_{2,i}\bar{g}_{1,i})
\delta(k\bar{g}_{2,f}k^{-1}\bar{g}_{1,i})
\delta_{\t_{1}}(\bar{g}_{2,f})\delta_{\t_{2}}(\bar{g}_{1,f}).
\ee
The effect of the exchange on the scalar product of two spin network
is then
\be
\bra\Phi_f|\Phi_i\ket
=\int \left(\prod_p dg_p{\delta_{\t_p}}(g_p)\right)
\bar{\Phi}_f(g_1g_2g_1^{-1},g_1,\cdots g_n){\Phi}_i(g_1,g_2,\cdots,g_n)
\delta(g_1\cdots g_n).
\ee
The effect of the exchange of particles $1,2$ therefore results
in the following operator acting on the physical Hilbert space
\be
{\cal R}_{\t_1,\t_2}\Phi_P(u_1,u_2,\cdots, u_{n-1})
=\Phi_P(u_1h_{\t_1}u_1^{-1}u_2,u_1,\cdots, u_{n-1}).
\ee
We recognize the action of the R-matrix of $D_\kappa(\SU(2))$
acting on the vector space ${\cal H}_{\t_1,0}\otimes {\cal H}_{\t_2,0}$
\cite{Koornwinder:1998xg}.

The effect of the exchange of two spinning particles
also result in the action of the
$\cal R$ matrix of $D_\kappa(\SU(2))$ on the physical Hilbert space
\be
{\cal R}_{(\t_1,s_1),(\t_2,s_2)}\Phi_P(u_1,u_2,\cdots, u_{n-1})
=\Phi_P(u_1h_{\t_1}u_1^{-1}u_2,u_1,\cdots, u_{n-1}),
\ee
where $\phi_P \in \otimes_v {\cal H}_{s_v}$.

In the appendix we present, starting from a triangulation, the computation of the
propagator in the case of a genus $g$ surface.
The result is given by the following propagator
\begin{equation}
G(a_i,b_i;a'_i,b'_i)=\int dk \d(\prod_{i=1}^g [a_i,b_i]) \prod_{i=1}^g
\d(a_i'ka_i^{-1}k^{-1})\d(b_i'kb_i^{-1}k^{-1}),
\end{equation}
where $[a,b]=aba^{-1}b^{-1}$ denotes the group commutator.
The general transition amplitude for a genus $g$ with $n$ punctures
 can be deduced and we recover the expected transition amplitude.
 \be\label{fullprop}
 \bra\Phi|\Phi\ket_{\Sigma_{g,n}}=
 \int\prod_{i=1}^{g} da_idb_i\prod_{p=1}^ndu_p
 |{\Phi}(a_i,b_i,u_ph_{\t_p}u_p^{-1})|^2
 \d(\prod_{i=1}^g [a_i,b_i]\prod_{p=1}^nu_ph_{\t_p}u_p^{-1}).
 \ee
It is interesting to compute  the vacua to vacua scalar product
\be
\bra 0|0 \ket_{\Sigma_{g,n}} = \sum_j \frac{1}{(d_j)^{2-2g-n}} \prod_{p=1}^{n}\frac{\chi_{j}(h_{\t_p})}{d_j},
\ee
where $d_j=2j+1$ is the dimension of the spin $j$ representation.
This formula gives the volume of the moduli space of flat $\SU(2)$ connections
\cite{Witten2}.

\subsection{Closed manifolds and finiteness of the modified PR model}\label{closem}
In the previous example we have seen that the Ponzano-Regge transition amplitudes
are explicitly finite after gauge fixing.
In the case of closed manifold this is not true in general.
Even after gauge fixing the
 Ponzano-Regge partition function is
not necessarily finite for closed manifold.
This is clearly seen if one tries to compute for instance
the partition function $Z(\Sigma_{g,n}\times S^1)$
which can be expressed as the trace of the propagator (\ref{fullprop}).
It should be so since such an expression is computing the dimension of the
physical Hilbert space which is of course infinite.
One natural question to ask is whether we can find
manifolds for which the gauge fixed Ponzano-Regge partition
function is finite.
One can  find necessary conditions, for example
 if a closed manifold $M$ contains a non contractible torus then the partition function
is infinite even after gauge fixing.

Lets consider, for instance, a mapping
Tori $\Sigma_f =\Sigma\times I/\sim_f$ where $f$ is a mapping class group
element ( a non trivial diffeomorphism of $\Sigma_{g,0}$) and
the equivalence relation is $x \times \{0\} = f(x) \times \{1\}$.
It is then known that if $\Sigma_f$ is hyperbolic
then the manifold is atoroidal.
One would expect the gauge fixed Ponzano-Regge partition function to be finite in that case,
but we haven't checked.
Also the evaluation (\ref{mathgf}) of the partition function
shows that the invariant is defined by counting the number of flat $SU(2)$
connections. This is very similar to the original definition of the Casson-Walker
invariant \cite{CasWaker} and we therefore expect the gauge fixed PR partition function to
be finite for integral homology spheres.

%%%%%%%%%%%%%%%%%%%%%%%%%%%%%%%%%%%%%%%%%%%%%%%%%%%%%%%%%%%%%%%%%%%%%%
%%%%%%%%%%%%%%%%%%%%%%%%%%%%%%%%%%%%%%%%%%%%%%%%%%%%%%%%%%%%%%%%%%%%%%
%%%%%%%%%%%%%%%%%%%%%%%%%%%%%%%%%%%%%%%%%%%%%%%%%%%%%%%%%%%%%%%%%%%%%%
%%%%%%%%%%%%%%%%%%%%%%%%%%%%%%%%%%%%%%%%%%%%%%%%%%%%%%%%%%%%%%%%%%%%%%
%%%%%%%%%%%%%%%%%%%%%%%%%%%%%%%%%%%%%%%%%%%%%%%%%%%%%%%%%%%%%%%%%%%%%%
%%%%%%%%%%%%%%%%%%%%%%%%%%%%%%%%%%%%%%%%%%%%%%%%%%%%%%%%%%%%%%%%%%%%%%
%%%%%%%%%%%%%%%%%%%%%%%%%%%%%%%%%%%%%%%%%%%%%%%%%%%%%%%%%%%%%%%%%%%%%%
%%%%%%%%%%%%%%%%%%%%%%%%%%%%%%%%%%%%%%%%%%%%%%%%%%%%%%%%%%%%%%%%%%%%%%

\section{Conclusion}

In this paper, we have given a complete treatment of the quantization
of 3 dimensional Euclidean gravity in the spin foam language
including an analysis of local Poincar\'e invariance, finiteness, topological invariance
and the insertion of massive spinning particles.
We have given a general prescription allowing us to compute
quantum transition amplitudes with interacting particles.
We have introduced the notion of  particle graphs functionals which generalize the notion
of Feynman graphs for theories coupled to three dimensional gravity.
We have sketched a new hamiltonian treatment of 3d gravity coupled to spinning particles
and showed that our amplitude prescription computes its physical scalar product.
We have presented the link between the spin foam quantization and the combinatorial quantization of
Chern-Simons.

We feel that our work opens the way to many new developements.
First, the treatment we have presented of the gauge fixing of the symmetries
should allow us to tackle the more challenging problem of the spin foam quantization
of Lorentzian 3d gravity with particles.
Then, our treatment of particle insertions in spin foam shows a clear link between
spin foams and Feynman graphs which need to be better understood and eventually generalized to
higher dimensional gravity.
It would also be very interesting to construct explicitly the field theory reproducing the amplitude
we have given. Our work gives many hints towards the answer.
The spin foam formalism we have developed here to include particles should be naturally extendable
 to the case where a cosmological constant is present.
One also needs to develop a better understanding concerning the insertion of time
in the quantum amplitude, and the corresponding semi-classical interpretation.
Eventually, we have to see wether the structures we have introduced here to include matter
can be exported for the study of 4d quantum gravity amplitudes.

\begin{acknowledgments}
We would like to thank Etera Livine for his help and  numerous key discussions.
We would like to thank Jose Zapata for discussion on the hamiltonian framework,
 Karim Noui and Alejandro Perez for sharing  with us some of their results.
We are grateful to Bernd Schroers, Justin Roberts and John Barrett for the organisation of
the ICMS Edinburgh workshop (29 June- 5 July 2003) where some of our results were presented.
D. L. has benefited from a Menrt and Eurodoc grants and the hospitality of PI.
\end{acknowledgments}

\appendix

\section{Notations on $\SU(2)$} \label{app:su2}
We use the following notations for $\SU(2)$. The elements of the
Cartan subgroup $H=U(1)$ are represented using
\begin{equation}\label{ht}
h_\f=\mm{e^{i{\f}}}{0}{0}{e^{-i {\f}}}.
\end{equation}
The Weyl group consists of the identity and the reflection $h_\phi \rightarrow h_{-\phi}$.
Let us choose a (non necessarily normalized) Haar measure on the group, the integration
over the group can be written as an integration over the conjugacy classes using the Weyl integration formula
\begin{equation}
\int_G dg f(g) = \int_H \frac{ \D(\t)^2}{|W|} \left(\int_{G/H} f(xh_{\t}x^{-1})dx\right) d\t
\end{equation}
where $H=U(1)$ denote the cartan subgroup,
 $\D(\t)=\sin\t$ and $|W|$ is the order of the Weyl group.

The representations of $\SU(2)$ are labelled by a half-integer $j$
and are realized on the spaces $V^j\sim\C^{2j+1}$. The matrix
elements of representations are given by the Wigner functions
$D^j_{mn}(g)$ satisfying the orthogonality property
\begin{equation}
\int  D^j_{mn}(g) \overline{D^{j'}_{m'n'}(g)} dg  =
\d_{j,j'}\frac{V_G}{d_j} \d_{m,m'} \d_{n,n'},
\end{equation}
$V_G$ being the volume of the group.
This relation can be written in terms of convolution product for characters
\be
\int_G  \overline{\chi_{j'}(g)} \chi_j(gx) dg =V_G \d_{j,j'}\frac{ \chi_j(x)}{d_j}.
\ee

We define the distribution $\delta_\phi(g)$ to be the distribution forcing
$g$ to be in the conjugacy class of $h_\phi$.
It is invariant under conjugation $\delta_\phi(g)= \delta_\phi(xgx^{-1})$
and normalized by
\be\label{dphi}
\int_G  \delta_\phi(g) f(g) dg =  \int_{G/H} f(xh_\phi x^{-1}) dx.
\ee
We can  write this distribution in terms of characters
\be
\delta_\phi(g)=\frac{1}{V_H} \sum_j \chi_j(h_\phi)\chi_j(g),
\ee
where $V_H= 2\pi$ is the volume of the Cartan subgroup.
The Weyl integration formula imply that
\be
\int_{H/W} d\phi \Delta^2(\phi) \delta_\phi(h_\theta) =1.
\ee
This means that we can relate this distribution to $\delta_\phi(\theta)$
the delta function on $H/W$,
\be\label{deltaHW}
  \delta_\phi(h_\theta)  =  \frac{\delta_\phi(\theta)}{\Delta^2(\phi)}.
\ee

\section{A delta function identity}\label{app:delta}

In this section we prove the following formula
\be \label{delta}
\int d^3x e^{i \tr(X g)} ( 1\pm \epsilon(g))= 8\pi \delta(\pm g),
\ee
where $X= x^i \sigma_i$ is in the Lie algebra, $g=\exp(i\theta {n}^i \sigma_i)$ is a $\SU (2)$ group element
with $\theta \in [0,\pi]$ and  $ n^in_i =1$; $\sigma^i$ are the Pauli matrices, $\tr{\sigma_i \sigma_j}=\delta_{ij}$,
 $ \epsilon(g)= \mathrm{sign(\cos\theta)} $, and $\delta(g)$ is the delta function on the group
with respect to the normalized Haar measure.
First we evaluate
\be
\int d^3x e^{i \tr(X g)}= (2\pi)^3 \delta^{(3)}(\sin\theta \vec{n}).
\ee
We can use the familiar identity $ \delta^{(3)}(\vec{X}) = \frac{1}{4\pi|X|^2} \delta(|X|)$ to write this evaluation as
\bea \label{deltaeq}
\frac{2\pi^2}{(\sin\theta)^2} \delta( |\sin\theta|)
&=& \frac{2\pi^2}{(\sin\theta)^2}\sum_{n\in Z} \delta( \theta -\pi n ) \frac{1}{|\cos \theta|},\\
&=& \frac{2\pi^2}{(\sin\theta)^2 }\sum_{n \in \Z} \left(\delta( \theta - \pi 2n ) +
\delta( \theta - \pi( 2n+1) )\right).
\eea
The normalized Haar measure on the group is given by
\be
dg = \frac{2}{\pi} d\theta (\sin\theta)^2 d^2n,
\ee
where $d^2 n$ is the normalized measure on $S^2$, therefore we can write
\bea
\delta(g)= \frac{\pi}{2(\sin\theta)^2} \sum_{n\in \Z}\delta( \theta - \pi 2n ), \\
\delta(-g)= \frac{\pi}{2(\sin\theta)^2} \sum_{n\in \Z} \delta( \theta - \pi (2n+1)).
\eea
Together with (\ref{deltaeq}) this proves that
\be
\int d^3x e^{i \tr(X g)} ={4\pi} \left(\delta(g) + \delta(-g)\right).
\ee
A similar computation shows that
\be
\int d^3x e^{i \tr(X g)}\epsilon(g) ={4\pi} \left(\delta(g) - \delta(-g)\right)
\ee
which proves eq(\ref{delta}).

\section{Explicit computations}\label{sec:computations}

In this appendix, we present explicit computations using the gauge
fixing procedure we introduced. In particular, we compute
partition functions for manifolds $\S_g\times S^1$ where $\S_g$ is
the genus $g$ surface, and transition amplitudes and propagator
for $\S_g\times I$. We present the explicit computations for the
genus 1 case and the results for the general genus, which can be
obtained in the same way.

\subsection{Partition function}

First, the results are the following : the gauge fixed partition
function for a manifold $\S_g\times S^1$ is given by
\begin{equation}
Z[\S_g\times S^1]=\int_{G^{2g+1}} dk d\u(a_1,b_1) \prod_{i=2}^g
da_i db_i\ \d(\prod_i^g a_ib_ia_i^{-1}b_i^{-1}) \prod_i^g
\d(a_ika_i^{-1}k^{-1})\d(b_ikb_i^{-1}k^{-1}).
\end{equation}

\begin{figure}[ht]
\includegraphics[width=6cm]{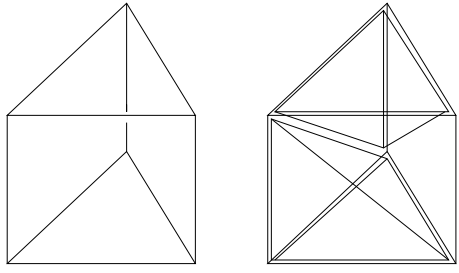}
\caption{Triangulation of a prism with 3 tetrahedra. The
tetrahedron owning past face is called $P$, the one owning future
face is called $F$, the intermediate one is called
$I$.}\label{fig:prisms}
\end{figure}

\begin{figure}[ht]
\begin{minipage}[t]{0.98\linewidth}
\begin{minipage}[t]{0.48\linewidth}
\includegraphics[width=0.47\linewidth]{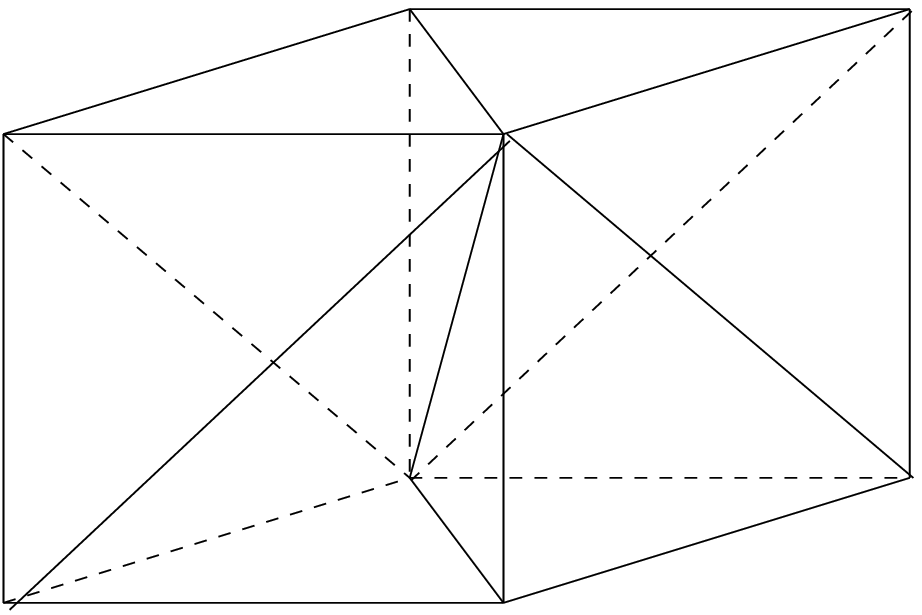}
\end{minipage}
\begin{minipage}[t]{0.48\linewidth}
\psfrag{f1}{$F_1$}\psfrag{i1}{$I_1$}\psfrag{p1}{$P_1$}
\psfrag{f2}{$F_2$}\psfrag{i2}{$I_2$}\psfrag{p2}{$P_2$}
\includegraphics[width=0.47\linewidth]{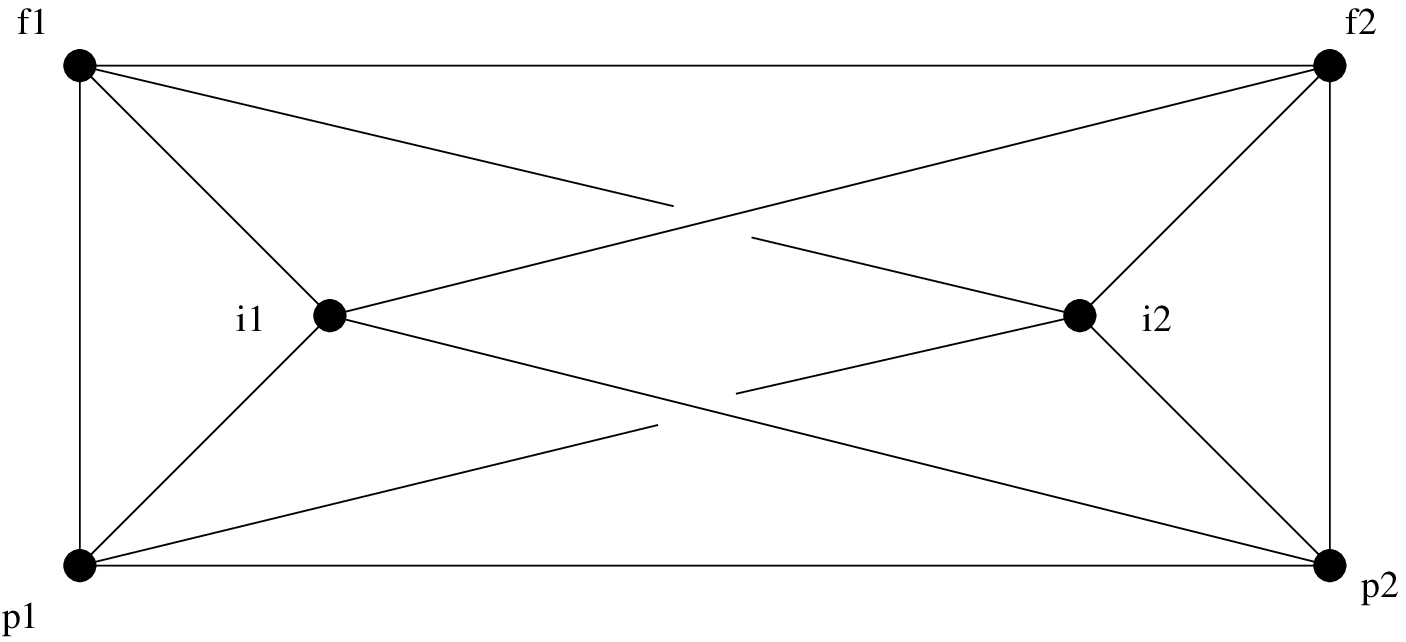}
\end{minipage}
\end{minipage}
\caption{Triangulation of $\S_1\times S^1$ and its dual
1-skeleton}\label{fig:triangS1}
\end{figure}

We give the explicit proof in the genus 1 case i.e $\S_1\times
S^1$ which is the 3 dimensional torus. A triangulation is obtained
by considering the triangulation of $\S_1$ with two triangles. It
induces a decomposition of $\S_1\times I$ into two prisms, each
prism can be triangulated with three tetrahedra (see figure
\ref{fig:prisms}). By identifying the past and future faces, we
obtained a triangulation of $\S_1\times S^1$ with 6 tetrahedra, 12
faces, 7 edges and 1 vertex (see figure \ref{fig:triangS1}). The 6
tetrahedra (dual vertices) are denoted $P_1,I_1,F_1,P_2,I_2,F_2$.

The non-gauge fixed partition function for $\S_1\times S^1$ takes
the form (\ref{eqn:discreteint-PI}) with 12 group elements and 7
$\d$-functions. The 7 dual faces corresponding to the 7 edges are
\begin{eqnarray}
&& F_1P_1I_2P_2F_2I_1, %
 F_2P_2I_1P_1F_1I_2, %
 I_1P_2P_1I_2F_1F_2,
P_1P_2I_2F_2F_1I_1,\\
&& P_1P_2I_2F_2F_1I_1, %
 F_2I_1P_1I_2, %
 P_2I_1F_1I_2, %
 F_1F_2P_2P_1,%
\end{eqnarray}

\begin{figure}[ht]
\psfrag{f1}{$F_1$}\psfrag{i1}{$I_1$}\psfrag{p1}{$P_1$}
\psfrag{f2}{$F_2$}\psfrag{i2}{$I_2$}\psfrag{p2}{$P_2$}
\includegraphics[width=5cm]{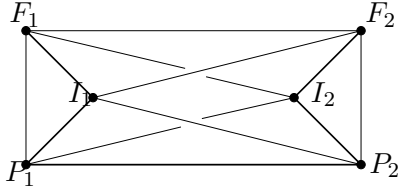} \caption{Gauge fixing on a maximal tree for the Lorentz symmetry on $\S_1\times S^1$ :
the thick edges are gauge fixed to identity}\label{fig:gaugeS1}
\end{figure}

The translational symmetry does not need to be gauge fixed since the
triangulation possess only one vertex. We only have to perform the
gauge fixing of the Lorentz symmetry. There are 6 dual vertices,
we choose to gauge fix to the identity the following 5 group
elements (see figure \ref{fig:gaugeS1}).
\begin{equation}
g_{I_1F_1},\ g_{I_1P_1},\ g_{I_2F_2},\ g_{I_2P_2},\ g_{P_1P_2}.
\end{equation}

\noindent After this gauge fixing, the partition function can be
written as
\begin{eqnarray}
&&Z[\S_1\times S^1]=\int
dg_{F_1F_2}dg_{P_1F_1}dg_{P_2F_2}dg_{I_1F_2}dg_{P_1I_2}
dg_{I_1P_2} dg_{F_1I_2} \nonumber\\
&&\d(g_{F_1P_1}g_{P_1I_2}g_{P_2F_2}g_{F_2I_1}) %
\d(g_{F_2P_2}g_{P_2I_1}g_{P_1F_1}g_{F_1I_2})  %
\d(g_{I_1P_2}g_{P_1I_2}g_{I_2F_1}g_{F_2I_1}) \nonumber \\ %
&&\d(g_{F_2I_1}g_{P_1I_2}) %
\d(g_{P_2I_1}g_{F_1I_2}) %
\d(g_{F_1F_2}g_{F_2P_2}g_{P_1F_1}) %
\d(g_{F_1F_2}).
\end{eqnarray}
Solving first the last $\d$ function gives $g_{F_1F_2}=1$, then the
three ones remaining on last line give
\begin{equation}
g_{I_1P_2}=a=g^{-1}_{I_2F_1}, \ g_{P_1I_2}=b=g^{-1}_{F_2I_1}, \
g_{P_1F_1}=k=g^{-1}_{F_2P_2}.
\end{equation}
This leads for the gauged fixed partition function to
\begin{equation}
Z[\S_1\times S^1]=\int_{G^3} da db dk\
\d(aba^{-1}b^{-1})\d(aka^{-1}k^{-1})\d(bkb^{-1}k^{-1}).
\end{equation}
Finally we use the measure $d\u(a,b)$ to take into account the
remaining gauge invariance at the last dual vertex.

The previous computation can be systematically generalized to the
genus $g$ case. We can triangulate $\S_g$ with $4g-2$ triangles.
This induces a triangulation of $\S_g\times S^1$ with $12g-6$
tetrahedra and still one vertex. The Lorentz gauge fixing
procedure can be systematically conducted along the same lines and
we obtained the general result announced.

\subsection{Propagator and transition amplitudes}

\begin{figure}[ht]
\includegraphics[height=4cm]{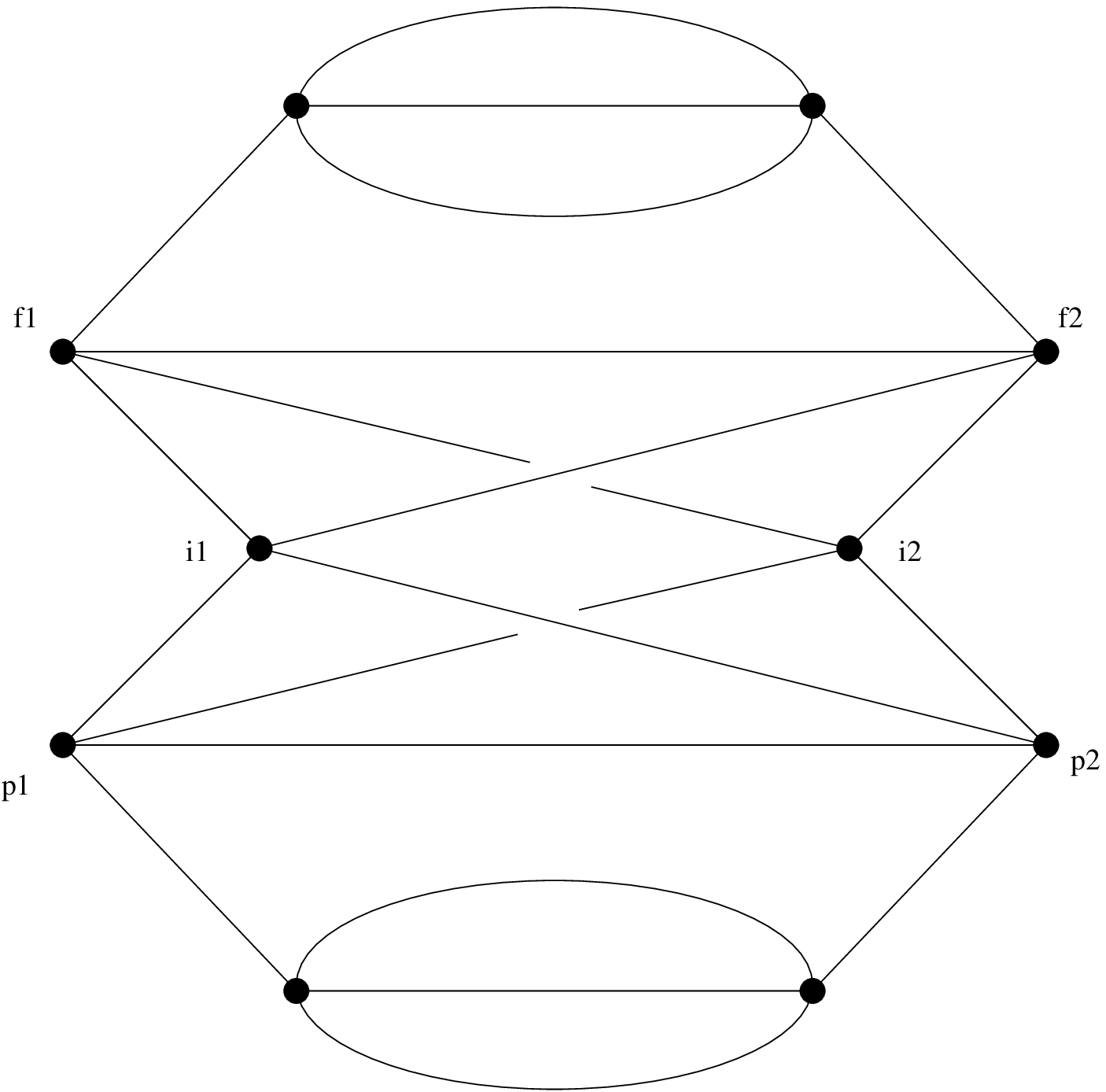}
\includegraphics[height=4cm]{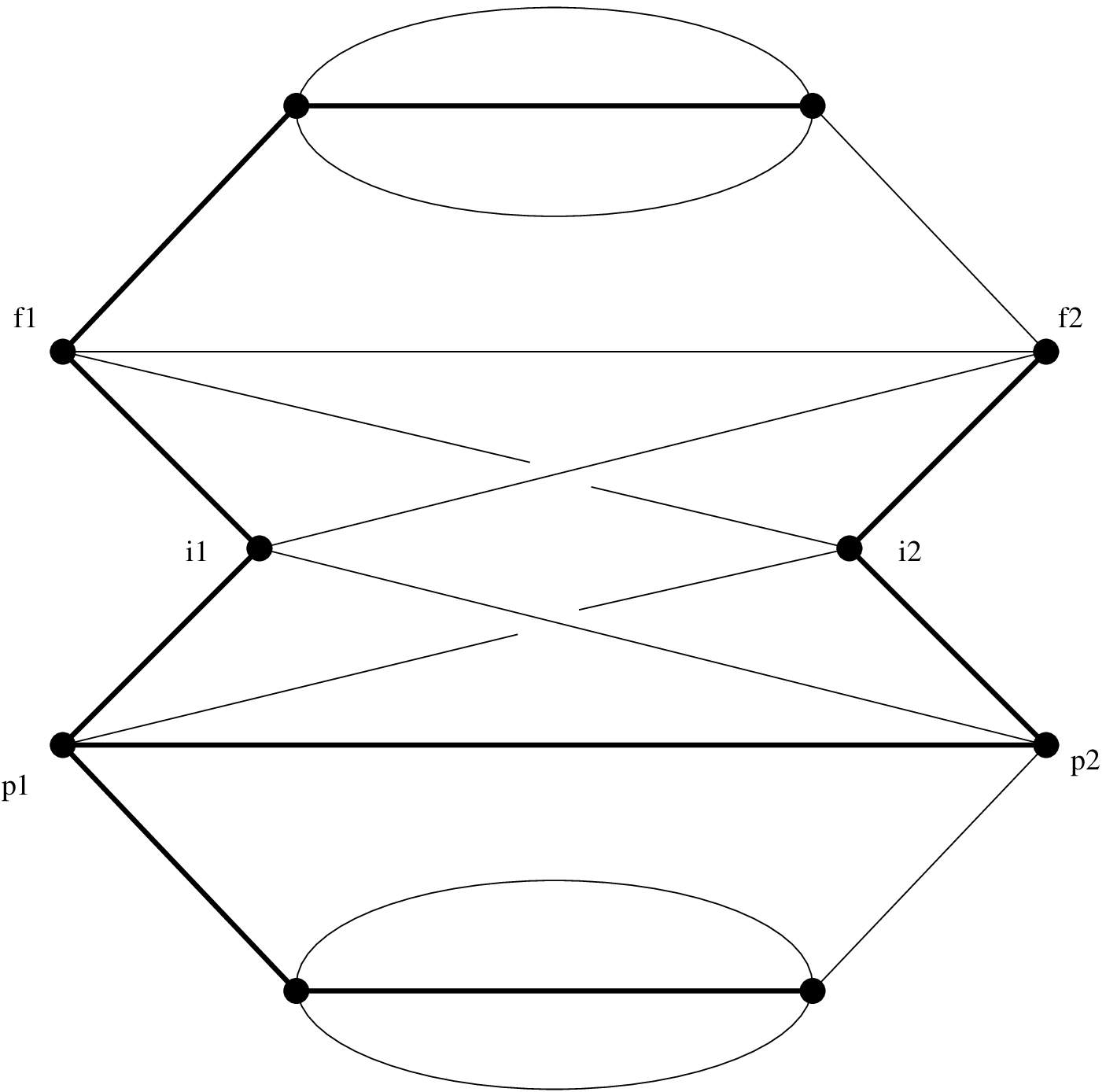}
\caption{Dual 1-skeleton of the triangulation of $\S_1\times I$
and maximal tree for the gauge
fixing}\label{fig:dualT2I}\label{fig:gaugeI}
\end{figure}

The propagator for genus $g$ case is given by
\begin{equation}
G(a_i,b_i;a'_i,b'_i)=\int dk \d(\prod_i a_ib_ia_i^{-1}b_i^{-1}) \prod_i
\d(a_i'ka_i^{-1}k^{-1})\d(b_i'kb_i^{-1}k^{-1}).
\end{equation}
Again we consider only the genus 1 case, i.e the case of the
manifold $\S_1\times I$, where the boundaries are a past and a
future surface $\S_1$. We consider the same way to triangulate
this manifold than before, with two triangles on each boundary and
6 tetrahedra (except that there is no more past/future
identification). We first consider the case of the propagator
between boundary connections. Each boundary carries three fixed
group elements representing the fixed boundary data, and we have
$14$ group elements living on the internal dual edges. There are 4
flatness conditions associated to internal edges of $\D$, and 6
flatness conditions around edges of the boundary (see figure
\ref{fig:dualT2I})

The dual graph is given by figure \ref{fig:dualT2I}. We now
perform a gauge fixing according to figure \ref{fig:gaugeI}.
The gauge fixing on the boundary spin-network leave only two group
elements. Writing explicitly the gauge fixed partition function
and solving the delta functions, the remaining propagator is
\begin{equation}
G(a,b;a',b')=\int dk \d(aba^{-1}b^{-1})\d(a'ka^{-1}k^{-1})\d(b'kb^{-1}k^{-1}).
\end{equation}
Recall that we use the measure $d\u(a,b)$ (see
(\ref{eqn:measureFL})) to fix the remaining gauge invariance at
the last vertex.
The gauge fixing procedure we proposed can be generalized in a
systematic way for the $\S_g\times I$ case. We obtain for the
propagator
\begin{equation}
G(a_i,b_i;a'_i,b'_i)=\int dk \, \d(\prod_i a_ib_ia_i^{-1}b_i^{-1}) \prod_i
\d(a_i'ka_i^{-1}k^{-1})\d(b_i'kb_i^{-1}k^{-1}).
\end{equation}

\end{document}